\documentclass[aps,prd,twocolumn,noshowpacs,nofootinbib]{revtex4}

\usepackage{amssymb} \usepackage{amsmath} \usepackage{graphicx}
\usepackage{epsfig,latexsym}

\RequirePackage{xspace} \allowdisplaybreaks
\usepackage{appendix}

\usepackage{color}

\begin{document}

\def\bef{\begin{figure}}
\def\eef{\end{figure}}

\newcommand{\nl}{\nonumber\\}

\newcommand{\ans}{ansatz }
\newcommand{\be}[1]{\begin{equation}\label{#1}}
\newcommand{\beq}{\begin{equation}}
\newcommand{\ee}{\end{equation}}
\newcommand{\beqn}[1]{\begin{eqnarray}\label{#1}}
\newcommand{\eeqn}{\end{eqnarray}}
\newcommand{\bd}{\begin{displaymath}}
\newcommand{\ed}{\end{displaymath}}
\newcommand{\mat}[4]{\left(\begin{array}{cc}{#1}&{#2}\\{#3}&{#4}
\end{array}\right)}
\newcommand{\matr}[9]{\left(\begin{array}{ccc}{#1}&{#2}&{#3}\\
{#4}&{#5}&{#6}\\{#7}&{#8}&{#9}\end{array}\right)}
\newcommand{\matrr}[6]{\left(\begin{array}{cc}{#1}&{#2}\\
{#3}&{#4}\\{#5}&{#6}\end{array}\right)}

\newcommand{\cvb}[3]{#1^{#2}_{#3}}
\def\lsim{\raise0.3ex\hbox{$\;<$\kern-0.75em\raise-1.1ex
e\hbox{$\sim\;$}}}
\def\gsim{\raise0.3ex\hbox{$\;>$\kern-0.75em\raise-1.1ex
\hbox{$\sim\;$}}}
\def\abs#1{\left| #1\right|}
\def\simlt{\mathrel{\lower2.5pt\vbox{\lineskip=0pt\baselineskip=0pt
           \hbox{$<$}\hbox{$\sim$}}}}
\def\simgt{\mathrel{\lower2.5pt\vbox{\lineskip=0pt\baselineskip=0pt
           \hbox{$>$}\hbox{$\sim$}}}}
\def\unity{{\hbox{1\kern-.8mm l}}}
\newcommand{\eps}{\varepsilon}
\def\ep{\epsilon}
\def\ga{\gamma}
\def\Ga{\Gamma}
\def\om{\omega}
\def\omp{{\omega^\prime}}
\def\Om{\Omega}
\def\la{\lambda}
\def\La{\Lambda}
\def\al{\alpha}
\newcommand{\ov}{\overline}
\renewcommand{\to}{\rightarrow}
\renewcommand{\vec}[1]{\mathbf{#1}}
\newcommand{\vect}[1]{\mbox{\boldmath$#1$}}
\def\tm{{\widetilde{m}}}
\def\mcirc{{\stackrel{o}{m}}}
\newcommand{\Dm}{\Delta m}
\newcommand{\dm}{\varepsilon}
\newcommand{\tanb}{\tan\beta}
\newcommand{\nbar}{\tilde{n}}
\newcommand\PM[1]{\begin{pmatrix}#1\end{pmatrix}}
\newcommand{\up}{\uparrow}
\newcommand{\down}{\downarrow}
\def\omE{\omega_{\rm Ter}}

%
%%%%%%%%%%     mauri    %%%%%%%%%%%%%%%%%%%%%%%%%%%%%%%%%

\newcommand{\Dsusy}{{susy \hspace{-9.4pt} \slash}\;}
\newcommand{\DCP}{{CP \hspace{-7.4pt} \slash}\;}
\newcommand{\mc}{\mathcal}
\newcommand{\gr}{\mathbf}
\renewcommand{\to}{\rightarrow}
\newcommand{\gtc}{\mathfrak}
\newcommand{\wh}{\widehat}
\newcommand{\br}{\langle}
\newcommand{\kt}{\rangle}

%%%%%%%%%%%%%%%%%%%%%%%%%%%%%%%%%%%%%%%%%%%%%%%%%%%%%%%%%%

\def\lsim{\mathrel{\mathop  {\hbox{\lower0.5ex\hbox{$\sim$}
\kern-0.8em\lower-0.7ex\hbox{$<$}}}}}
\def\gsim{\mathrel{\mathop  {\hbox{\lower0.5ex\hbox{$\sim$}
\kern-0.8em\lower-0.7ex\hbox{$>$}}}}}
%%%%%%%%%%%%%%%%%%%%%%%%%%%%%%%%%%

\def\nn{\\  \nonumber}
\def\de{\partial}
\def\brf{{\mathbf f}}
\def\bbf{\bar{\bf f}}
\def\bF{{\bf F}}
\def\bbF{\bar{\bf F}}
\def\bA{{\mathbf A}}
\def\bB{{\mathbf B}}
\def\bG{{\mathbf G}}
\def\bI{{\mathbf I}}
\def\bM{{\mathbf M}}
\def\bY{{\mathbf Y}}
\def\bX{{\mathbf X}}
\def\bS{{\mathbf S}}
\def\bb{{\mathbf b}}
\def\bh{{\mathbf h}}
\def\bg{{\mathbf g}}
\def\bla{{\mathbf \la}}
\def\bmu{\mathbf m }
\def\by{{\mathbf y}}
\def\bmu{\mbox{\boldmath $\mu$} }
\def\bsig{\mbox{\boldmath $\sigma$} }
\def\bunity{{\mathbf 1}}
\def\cA{{\cal A}}
\def\cB{{\cal B}}
\def\cC{{\cal C}}
\def\cD{{\cal D}}
\def\cF{{\cal F}}
\def\cG{{\cal G}}
\def\cH{{\cal H}}
\def\cI{{\cal I}}
\def\cL{{\cal L}}
\def\cN{{\cal N}}
\def\cM{{\cal M}}
\def\cO{{\cal O}}
\def\cR{{\cal R}}
\def\cS{{\cal S}}
\def\cT{{\cal T}}
\def\eV{{\rm eV}}

%%%%%%%%%%%%%%%%%%%%%%%%%%%%%%%%%%%%%%

\title{A geometric phase approach to quark confinement\\ 
from stochastic gauge-geometry flows}

\author{Torsten Asselmeyer-Maluga}
\email{torsten.asselmeyer-maluga@dlr.de}
\affiliation{German Aerospace Center (DLR), Sachsendamm 61, 10829, Germany, EU}

\author{Antonino Marcian\`o}
\email{marciano@fudan.edu.cn}
\affiliation{Center for Astronomy and Astrophysics, Center for Field Theory and Particle Physics, and Department of Physics, Fudan University, Shanghai 200438, China}
\affiliation{Laboratori Nazionali di Frascati INFN, Frascati (Rome), Italy, EU}
\affiliation{INFN sezione Roma Tor Vergata, I-00133 Rome, Italy, EU}

\author{Roman Pasechnik}
\email{roman.pasechnik@fysik.lu.se}
\affiliation{Department of Physics, Lund University, S\"olvegatan 14A S 223 62 Lund, Sweden, EU}

\author{Emanuele Zappala} 
\email{emanuelezappala@isu.edu}
\affiliation{Department of Mathematics and Statistics, Idaho State University\\
Physical Science Complex,  921 S. 8th Ave., Stop 8085, Pocatello, ID 83209, USA} 

\begin{abstract}
\noindent
We apply a stochastic version of the geometric (Ricci) flow, complemented with the stochastic flow of the gauge Yang--Mills sector, in order to seed the chromo-magnetic and chromo-electric vortices that source the area-law for QCD confinement. The area-law is the key signature of quark confinement in Yang--Mills gauge theories with a non-trivial center symmetry. In particular, chromo-magnetic vortices enclosed within the chromo-electric Wilson loops instantiate the area-law asymptotic behaviour of the Wilson loop vacuum expectation values. The stochastic gauge-geometry flow is responsible for the topology changes that induce the appearance of the vortices. When vortices vanish, due to topology changes in the manifolds associated to the hadronic ground states, the evaluation of the Wilson loop yields a dependence on the length of the path, hence reproducing the perimeter law of the hadronic (Higgs) phase of real QCD. Confinement, instead, is naturally achieved within this context as a by-product of the topology change of the manifold over which the dynamics of the Yang--Mills fields is defined. It is then provided by the Aharonov--Bohm effect induced by the concatenation of the compact chromo-electric and chromo-magnetic fluxes originated by the topology changes. The stochastic gauge-geometry flow naturally accomplishes a treatment of the emergence of the vortices and the generation of turbulence effects. Braiding and knotting, resulting from topology changes, namely stochastic fluctuations of the Einstein–Yang--Mills system, stabilize the chromo-magnetic vortices and dynamically induce, as a non-trivial topological feature, the chiral symmetry-breaking. Finally, we observe that dimensional transmutation for the Yang-Mills fields can be derived from the scaling property of the geometric part of the stochastic flow. Specifically, a relation that involves the infrared equilibrium limit of the Planck constant can be derived that yields the correct order of magnitude for $\Lambda_{\rm QCD}$.   
\end{abstract}

\maketitle

%%%%%%%%%%%%%%%%%%%%%%%%%%%%%%%%%%
\section{Introduction}\label{sec:intro}
\noindent 
%%%%%%%%%%%%%%%%%%%%%%%%%%%%%%%%%%
The underlying mechanism of confinement in quantum Yang--Mills theories, with the related understanding of the way quarks in Quantum Chromo Dynamics (QCD) are bound into hadrons, has been a hot topic of investigation for decades. The community is currently seeking to reach a deeper understanding beyond semiclassical explanations that involve the possible crucial role of chromo-electric strings and of magnetic monopoles' condensates, as developed in seminal papers by Mandelstam, 't Hooft and Polyakov  \cite{Mandelstam:1974pi,Zichichi:1975izv,Polyakov:1975rs} --- for a recent review of the underlined concepts and implications, see e.g.~Refs.~\cite{Greensite:2011zz,Greensite:2016pfc,Pasechnik:2021ncb} and references therein.

An insight about the relevance of topology for confinement was offered by 't Hooft in his renown paper \cite{tHooft:1977nqb}. Sets of operators that modify the gauge-topological structure of gauge field theories were considered, showing that physical effects nonetheless can be local. In $2+1$ dimensions, it was demonstrated that in the absence of a spontaneous breakdown of the local gauge symmetry, topological fields acquire a vacuum expectation value (vev) that spontaneously breaks their mutual global symmetry. This mechanism then induces permanent confinement of the quarks. In $3+1$ dimensional non-abelian gauge theories it was shown that topological operators and gauge field operators close an algebra that explains the occurrence of four possible phases: i) spontaneous breakdown induced by a Higgs field; ii) permanent confinement of gauge quantum numbers in the absence of the Higgs field, either explicit or composite; iii) confinement in presence of the Higgs mechanism; iv) an intermediate phase, presumably a critical point, holding massless particles. The phases i) and ii) were shown, for a simple realization of the algebra, to be connected by dual transformations. Phase ii) was called superinsulating, as dual to the super-conductive phase. Along these lines, Diamantini, Trautenberger and Vinokur proposed in \cite{Diamantini:2018mjg} that the superinsulating state, characterized by an infinite resistance, may emerge on the insulating side of the superconductor-insulator transition in superconducting films, providing a realization of confinement with direct experimental access.\\

Topology entered hitherto without involving geometric considerations, but solely referring to the properties of the vacuum states, intended as background states. Here, we propose a program towards a dynamical description of confinement in QCD that is rooted in geometric topology, geometry, and stochastic quantization techniques. The theoretical framework we develop intertwines among concepts of topology and out-of-equilibrium physics. In our framework, the quark confinement phenomenon is described as a geometric phase effect that emerges dynamically, according to a stochastic flux along thermal time that drives the system toward relaxation. The stochastic flux implements a scale transformation in the Renormalization Group (RG) sense, and is responsible for the change of topology of the manifold associated to the QCD ground-state under scrutiny. Nucleons, as well as mesons and more complex composite resonances, are indeed seen as low-curvature manifolds of increasingly high topology classes characterized by the stochastic flow.\\

The change of topology in the manifolds, on the other hand, is interpreted in the dual picture as the generation of flux-lines of the gauge connection (gluon field). These flux-lines source Wilson loops and center vortices ultimately reproducing the asymptotic area law of the Wilson-loop vevs. The phase transition between the perimeter- and the area-law regimes shall therefore be addressed in terms of the topology changes that are induced by the stochastic geometric flow. This latter is coupled to the stochastic RG flow of the gauge field through the out-of-equilibrium dynamics in the thermal time. The stochastic gauge-geometry flux then drives the relaxation of the system toward equilibrium configurations, these latter being characterized by the Einstein--Yang-Mills equations.\\

The strategy we propose involves the following three stages:
\begin{itemize}
    \item i) 
    Addressing stochastic gradient flows, either geometry or gauge/matter, in light of the changes of topology they induce on the manifolds by originating singularities; 
    %Establishing the framework of topological RG (TRG), a version of the Ricci flow (RF) tailored to topological quantum field theories (TQFTs). While the RF implements rescaling of topologies through conformal transformations, TRG complements the RG with changes of topology;
    
    \item ii) Advocating the stochastic quantization method to leverage stochastic gradient flows. The stochastic geometric flow, or stochastic Ricci flow, is then extended to the Wilsonian formulation of Yang--Mills theories. We furthermore consider the  interaction among the gauge sector and the gravitational field, since this affects the topology of the electric and magnetic Yang--Mills configurations;
    
    \item iii) Developing a geometric phase approach to confinement. The geometric curvature is not relevant to the purpose of these studies. What really matters is topology: the reason to consider the interaction with gravity is therefore finding the dynamics that induces through changes of topology, the generation of vortices in the Yang--Mills gauge sector, which finally source the area-law scaling.
\end{itemize}

Achieving these steps realizes a novel proposal for a dynamical description of QCD confinement, a non-perturbative phenomenon responsible for a plethora of quark-gluon and hadron physics implications in the infrared regime of QCD.

The plan of the paper is the following. In Sect.~\ref{sec:conf-mech} we provide an overview of the center vortex mechanism of confinement, serving as a baseline for further considerations. In Sect.~\ref{sec:theory} we outline the (stochastic) gauge-geometry flow formulation and its implication for the dynamics of confinement. In Sect.~\ref{sec:out-of-eq} we elaborate on the important details of the out-of-equilibrium dynamics in thermal time and on the stochastic gradient flow. The formulation of the Wilsonian approach to QCD and dynamical stochastic evolution of topology in the context of confinement is provided in Sect.~\ref{sec:wilson}. Then, in Sect.~\ref{sec:proposal} we present a novel picture of QCD confinement dynamics resorting to the geometric (topology) fluctuations due to the stochastic gauge-geometry flow. Conclusions and outlook follow in Sect.~\ref{sec:conclusions}. Finally, in Appendix~\ref{sec:topology} we provide details of the topology and geometry of the $3$-manifold relevant for our formulation, while in Appendix~\ref{Casson} we review the embedding of Casson handles in the presence of quadratic differentials and vertical foliations. 

%%%%%%%%%%%%%%%%%%%%%%%%%%%%%%%%%%%%%%%%%%%%%%%%%%%%%%%%%%%%%%%%%%%%%%
\section{Confinement and center vortices}\label{sec:conf-mech}
\noindent 
%%%%%%%%%%%%%%%%%%%%%%%%%%%%%%%%%%%%%%%%%%%%%%%%%%%%%%%%%%%%%%%%%%%%%%
In this section, we overview the basics of the vortex mechanism of QCD confinement \cite{Engelhardt:1998wu} that features a number of numerical evidences as summarized in Ref.~\cite{Greensite:2016pfc}. A detailed review of this and other mechanisms of confinement is beyond the purpose of this paper and can be found e.g.~in Refs.~\cite{Greensite:2011zz,Pasechnik:2021ncb} and references therein. Therefore, here we only provide a brief introduction into the underlined concepts relevant for our subsequent formulation, pointing to the literature where further details can be found.

%%%%%%%%%%%%%%%%%%%%%%%%%%%%%%%
\subsection*{Center symmetry}
%%%%%%%%%%%%%%%%%%%%%%%%%%%%%%%
\noindent 
The Elitzur's theorem \cite{Elitzur:1975im} ensures that the phases of a gauge theory cannot be distinguished in terms of the breakdown of local gauge symmetries. Therefore, the breakdown of additional global symmetries must be considered that enables to identify gauge phases also in presence of a local order parameter --- this is indeed the case of the $\mathbb{Z}_2$ symmetry in the Ising model. At this point, we remind that fixing a covariant gauge connection does not fix completely the gauge freedom, leaving instead remnant symmetries that evade the Elitzur's theorem and may therefore undergo spontaneous symmetry breaking. This is the case of the Kugo-Ojima condition \cite{Kugo:1995km,Kugo:1979gm}, a suitable confinement criterion that requires the full residual symmetry in the Landau gauge $\partial^\mu A_\mu^a=0$ to remain unbroken, hence implying the vanishing in any physical state $\psi$ of the expectation value of $\langle \psi | Q_a |\psi \rangle$, the color charge operator. The global and space-time dependent parts of this full residual gauge symmetry with respect to the gauge transformation $A_\mu \rightarrow G A_\mu G^\dagger$ in the Landau gauge acquires the form \cite{Hata:1981nd,Hata:1983cs} 
\begin{eqnarray}\label{nax}
G(x)&=& \exp\left(\frac{\imath}{2} \Xi^a(x) \sigma^a \right)\,, \\   \Xi^a(x)  &=& \epsilon^a_\mu x^\mu -g \frac{1}{\partial^2 }(A_\mu \times \epsilon^\mu )^a + \mathcal{O}(g^2) 
\,,
\end{eqnarray}
with $\epsilon^a_\mu$ arbitrary parameters and $g$ the $SU(2)$ gauge coupling. In addition to \eqref{nax}, a further criterion must be fulfilled for confinement: another part of the residual gauge symmetry that is space-time independent remains unbroken.\\

The separation of phases is characterized by non-analytic boundaries between magnetic order and disorder configurations, which are associated with (either spontaneous or dynamical in thermal time) breaking of global center symmetries in Yang--Mills gauge groups. Specifically, a center symmetry is a subset of gauge group elements that commute with all the elements of the gauge group --- e.g. $\mathbb{Z}_N$ for $SU(N)$, where the group elements are $\exp 2 \pi \imath n/N$, with $n=0,\dots N-1$. The representations of $\mathbb{Z}_N$ individuate $N$ possible subsets, known as $N$-alities, belonging to infinite numbers of possible representations of $SU(N)$. A given $N$-ality $k$ corresponds to the number of boxes of Young tableau mod $N$. Therefore, $N$-alities represent how a given representation transforms under a given center symmetry group. For instance, for representations $R$ of $SU(N)$ group elements $g$, i.e. $R(g)$, the action of the center symmetry is expressed as $R(zg)=z^k\, R(g)$, with $z$ being an element of $\mathbb{Z}_N$ and $k$ the $N$-ality class. The Yang--Mills action 
\begin{equation}
 \mathcal{S}_{\rm W} = - \frac{1}{4 g^2} {\rm Tr}  [W_{\circ  \gamma}(A)+W_{\circ \gamma}^\dagger(A)] 
\end{equation}
is written in terms of the super-trace of the Wilson loops $W_{\circ \gamma}(A)$. These are holonomy operators $H_{\gamma}(A)$, depending on the gauge-field $A$, that are traced with respect to the representation indices and supported on loops $\circ \gamma$, the closure of the curves $\gamma$.\\

The singular gauge transformation, holding implicit dependence on the thermal time, applied to $H_\gamma(A)$ reads 
\begin{equation}
H_{\gamma(\tau)}(A) \rightarrow G_{\gamma(\tau_s)} H_{\gamma(\tau_0)}(A)G_{\gamma(\tau_t)}^\dagger\,,
\end{equation}
where $\tau_{s,t}$ are the thermal times at the source ($s$) and the target ($t$), and the periodicity condition
\begin{equation}
    G_{\circ \gamma}=z^*G_{\circ \gamma }
\end{equation}
holds up to a center symmetry transformation. A particular case of this transformation applied to $H_\gamma(A)$ is
\begin{equation}
H_{\gamma(\tau_0)}(A) \rightarrow  z \, H_{\gamma(\tau_0)}(A)\,, \quad {\rm with} \quad z\in \mathbb{Z}_N\,,
\end{equation}
with $\tau=\tau_0$ selecting a fixed thermal time-slice. This transformation, in the asymptotic thermal time limit, corresponds to 
\begin{equation} \label{NAX}
A_\mu(x) \rightarrow G(x)  A_\mu(x) G^\dagger(x) - \frac{\imath}{g} G(x) \partial_{\mu}  G^\dagger(x)\,. 
\end{equation}
The transformation \eqref{NAX} realizes at equilibrium an ``almost'' gauge symmetry, but only for $\mu=0$, and only at the base point of the loop, where the second term (turned into a $\delta$-function) is dropped --- hence, the notion of a ``singular'' transformation. Such a transformation does not leave the action invariant, but has a special role to be elaborated upon more in what follows.\\

Unbroken center symmetry in the pure Yang--Mills theory hence provides an important criterion for confinement, which is associated with an area-law fall-off of large Wilson loop vevs, effectively indicating the emergence of a magnetic disorder state. In our picture, which will be detailed in what follows, the out-of-equilibrium dynamics induces the relaxation of the system from magnetic order to magnetic disorder configurations, hence from the deconfined phase to a confined one.
Indeed, a magnetically disordered configuration implies a linear growth of the potential of static quarks asymptotically with the distance between them, thus yielding a linear string potential.\\

As the action appears to be non-invariant under singular gauge transformations, consequently, a loop of magnetic flux with a singularity is effectively induced by such a transformation. Furthermore, we conjecture that a singular gauge transformation is tightly connected to the out-of-equilibrium dynamics, being associated with the creation of thin center vortices in the gauge-field system.

%%%%%%%%%%%%%%%%%%%%%%%%%%%%%%%
\subsection*{Geometric phase}
%%%%%%%%%%%%%%%%%%%%%%%%%%%%%%%
\noindent 
A well-known example is provided by the abelian $U(1)$ case, whose closed holonomy on a spacelike loop $\circ \gamma$ provides a geometric phase 
\begin{equation}
W_{\circ \gamma}= e^{\imath e \Phi_B}\,,
\end{equation}
with $\Phi_B$ magnetic flux through $\circ \gamma$ and $e$ electric charge. When the loops wind around a solenoid oriented along the $z$-axis, one may find configurations corresponding to non-vanishing magnetic fluxes even for zero magnetic fields along $\circ \gamma$. This latter can be viewed as a result of the singular gauge transformation applied to a vanishing gauge field with a discontinuous $G(x)$. In cylindrical coordinates $\{r,\theta, z, t\}$ this entails 
a discontinuity in $\theta$ for $r>0$, namely,
\begin{equation}
W_{\circ \gamma} \rightarrow  e^{\pm \imath e \Phi_B}  W_{\circ \gamma} \,,
\end{equation}
with $\exp {\pm \imath e \Phi_B}$ a $U(1)$ group element and $\pm$ depending on the orientation of $\circ \gamma$ and corresponding to the production along the $z$-axis of a singular line of magnetic flux. The winding number is defined as the number of revolutions around a fixed point in two space dimensions. This can be also generalised in three space dimensions through the so-called linking number, specifying how many times the two loops wind around each other.\\

For a $SU(N)$ Yang--Mills theory in $d$ dimensions the geometric phase is replaced by a center-group element of $\mathbb{Z}_N$, namely
\begin{equation}
G_{\gamma(\tau_s)}=z G_{\gamma(\tau_t) }\,, \qquad W_{\circ \gamma} \rightarrow (z^*)^l W_{\circ \gamma}\,,    
\end{equation}
with $l$ linking number and $\circ \gamma$ a space-like Wilson-loop,  topologically linked to $(d-2)$-dimensional thin vortices. The quanta of these non-abelian magnetic fluxes are referred to as thin center vortices. The regularization of the singular color-magnetic field is then achieved by smearing out the vortices in directions transverse to the two-dimensional hypersurfaces. These are known as thick center vortices \cite{Greensite:2011zz}.

%%%%%%%%%%%%%%%%%%%%%%%%%%%%%%%
\subsection*{'t Hooft loop}
%%%%%%%%%%%%%%%%%%%%%%%%%%%%%%%
\noindent 
The 't Hooft loop can be introduced as the operator $B_{\circ \gamma}$ that creates thin center vortices at the base point $x_0=x(\tau_0)$ of a loop $\alpha\equiv \circ \gamma$, embedded in the base manifold $M$ of the principal bundle of the gauge theory. If two loops $\alpha$ and $\alpha'$ on a three-dimensional hypersurface are topologically linked, with linking number $l=1$, then by our previous discussions the following relation holds: 
\begin{equation}
B_\alpha W_{\alpha'}= z W_{\alpha'} B_\alpha\,, \qquad z\in \mathbb{Z}_N\,.
\end{equation}
The vev of a Wilson loop $W_\alpha$, which we denote as ${\cal W}_\alpha \equiv \langle W_\alpha\rangle$, and of a 't Hooft loop, namely $\langle B_\alpha\rangle$, in their asymptotic behaviour, either satisfy a perimeter-law or an area-law, but never simultaneously, as shown in \cite{tHooft:1977nqb}.

For $a$ and $b$ positive and real numbers, a confined (magnetically disordered) phase is achieved when
\begin{equation}
\mathcal{W}_\alpha \sim e^{-a A_\alpha} \quad \leftrightarrow \quad \langle B_\alpha \rangle  \sim e^{-b P_\alpha} \,,
\end{equation}
while a deconfined phase is enforced, for $a'$ and $b'$ positive and real numbers, when 
\begin{equation}
\mathcal{W}_\alpha \sim e^{-a' P_\alpha} \quad \leftrightarrow \quad \langle B_\alpha \rangle  \sim e^{-b' A_\alpha} \,,
\end{equation}
where with obvious notation $A$ and $P$ refer, respectively, to the area and the perimeter of the corresponding loop. 
The latter case implies a spontaneously broken center symmetry, entailing a magnetically ordered phase. Therefore, the Wilson loops and the 't Hooft loops can be considered as dual to one another: the former ones create loops of color electric fluxes, while the latter create closed loops of color-magnetic flux (thin center vortices), both at fixed thermal time $\tau$. 

%%%%%%%%%%%%%%%%%%%%%%%%%%%%%%%
\subsection*{Wilson loops and characterization of confinement}
%%%%%%%%%%%%%%%%%%%%%%%%%%%%%%%
\noindent
The Wilson loops operators, as closed holonomies, are represented as the path-ordered (typically, denoted as ``$\mathcal{P}$'') exponential of the gauge connection $A_\mu$, namely,
\begin{equation}
W_\alpha = \mathcal{P} \,\exp \left[ \imath g \oint
dx^\mu A_\mu (x) \right]\,.
\end{equation}
The path contour that is customarily picked is a rectangular time-like loop that describes the creation, the propagation and finally the annihilation of a pair of static quark and anti-quark that are placed at space-like base-points of the loop. In Euclidean space, this can be obtained, for instance, by combining operators that create particle-antiparticle pairs in a color singlet state at a given time $T_E$ and separation $R$, i.e. 
\begin{equation}
\mathcal{C}(T_E)= \psi^\dagger(0,T_E) H_{\gamma_R} \psi(R,T_E) \,, 
\end{equation}
where the holonomy of the gauge connection $H_{\gamma_R}$ along the line $\gamma_R$, from $0$ to $R$ at a fixed time $T_E$, creates a color-electric flux tube (string) among the charges. Neglecting the mass term of particles in the heavy-mass limit, the action for the gauge-matter system in the Euclidean lattice formulation is customarily written as 
\begin{eqnarray}
\mathcal{S}=&-&\frac{\beta}{N} \sum_p {\rm Tr} \left[ U_\alpha(A) + \mathrm{c.c.} \right]  \nonumber \\
&+& \delta \sum \limits_{x,\mu} \left[  \psi^\dagger (x) H_{\gamma_\mu}(x) \psi(x+\hat{\mu}) + \mathrm{c.c.} \right] \,,
\end{eqnarray}
with $\beta$ and $\delta$ lattice couplings, $\hat{\mu}$ denoting a generic direction in the Euclidean space, $H_{\gamma_\mu}(x)$ representing the holonomy along paths $\gamma_\mu$, oriented along generic directions $\mu$, which close a plaquette $p$ around a base point $x$. In the heavy matter fields limit an infinite energy would be required to pull pairs of particles and antiparticles away from the vacuum, placing them on mass-shell. A similar amount of energy would be also required to bind pairs to sources and screen their charges, hence stretching the flux tube (of strings) to an infinite length before they break apart, an {\it ad absurdum} argument that supports the picture of confinement of the quantum vacuum. \\

Integrating out the matter fields $\psi$ in the functional integral while considering the vev of the product operator $\mathcal{C}(T_E)^\dagger \mathcal{C}(0)$ is equivalent, in the heavy mass limit, to considering $\mathcal{W}_\alpha(R,T_E)$, the vev of the Wilson loop, namely
\begin{equation}
    \langle \mathcal{C}(T_E)^\dagger \mathcal{C}(0) \rangle \sim \mathcal{W}_\alpha(R,T_E)\,,
\end{equation}
with 
\begin{equation}
\mathcal{W}_\alpha(R,T_E)=\langle {\rm Tr}[H_{\gamma} H_{\gamma'} \dots  ]_{\alpha} \rangle \equiv \chi_\alpha[H(R, T_E)]\,,
\end{equation}
where $\chi$ denotes the trace of the representation of gauge group-elements.\\

In the operator formalism \cite{Greensite:2011zz} one can derive the asymptotic $T_E\rightarrow \infty$ limit 
\begin{equation}
   \langle \mathcal{C}(T_E)^\dagger \mathcal{C}(0) \rangle  \propto \sum_n |c_n|^2 e^{-\Delta E_n T_E}
   \sim e^{-\Delta E_{\rm min} \, T_E} \,,
\end{equation}
with $\Delta E_n$ the energy of the $n$-th excited states above the vacuum, and $\Delta E_{\rm min}$ the minimum among them, which individuates a dominant term contribution to the vev. The energy difference between two static charges, namely the interaction static potential $V(R)$, corresponds precisely to $\Delta E_{\rm min}$. Therefore, the asymptotic limit of the Wilson loop along the time-like (counterclockwise) rectangular path $\alpha=R\times T_E$, is found to be characterized by the potential $V(R)$.

If we now use the Stokes theorem, we can characterize different types of fall-off for the Wilson loops, hence distinguishing among confined and de-confined phases. For the paradigmatic case of $U(1)$, considering a planar and non self-intersecting Wilson loop one can find 
\begin{equation}
W_\alpha=\exp \left[\imath e \oint_\alpha dx^k A_k(x) \right] = \exp \left[\imath e \oint_\alpha dS_\alpha F_{ij}(x) \right]\,,
\end{equation}
where the integration individuates the magnetic flux over the minimal area enclosed by a large Wilson loop. The additive properties of a loop enables to decompose it into a product of smaller loops, picked in a way that neighbor loops run in opposite directions to one another, i.e.
\begin{equation}
W_\alpha=\prod\limits_{i=1,n} H_{\gamma_i}\,.   
\end{equation}
Magnetic disorder is accomplished when magnetic fluxes through the smaller loops of the decomposition (i.e. plaquette variables) are completely uncorrelated and their vevs factorize, i.e.
\begin{equation}
 \mathcal{W}_\alpha  =   \prod\limits_{i=1,n} \langle H_{\gamma_i} \rangle = \exp \left[ -\sigma_\gamma A_\alpha \right]\,,
\end{equation}
where
\begin{equation}
\sigma_\gamma = - \frac{\ln \langle H_\gamma \rangle }{A'} \,,
\end{equation}
and $A$ and $A'$ are the larger and smaller Wilson loop areas, respectively.\\ 

Within the absence of light matter fields, in principle able to screen the color charge of massive sources, and considering a rectangular Wilson loop supported over the path $\alpha=R\times T_E$, magnetically disordered states can be shown to be characterized by a linear growth of the interaction potential in the distance $R$ between the asymptotically static charges, namely,
\begin{equation}
V(R)=\sigma R + 2 V_0\,,
\end{equation}
which is the potential of a linear string of tension $\sigma$, in a given representation that does not depend on the sub-loop area $A'$, complemented with a self-energy contribution $V_0$. The area law is then recovered for $T_E\gg R$ for a generic counter that encloses a large minimal area $A_\alpha$, i.e.
\begin{equation}
    \mathcal{W}_{\alpha} \sim \exp \left[ -\sigma A_\alpha  - V_0 P_\alpha \right]\,,
\end{equation}
with $P_\alpha$ denoting the perimeter of the loop $\alpha$ in the limit $T_E\gg R$.\\

In general, in two space-dimensions one can show for any gauge group that only the magnetically disordered phase is achieved and that the area-law fall-off of the Wilson loop is recovered, due to the absence of Bianchi constraints on the components of the strength tensor of the gauge field \cite{Halpern:1978ik}. On space hypersurfaces of higher than two dimensions, the Bianchi identities correlate field strength values at neighbor sites, so that they cannot fluctuate independently from one point to another \cite{Batrouni:1984rb}. The absence of correlation among the arbitrary-size plaquettes in which a Wilson loop is decomposed is indeed at the onset of the area-law relation. On space-hypersurfaces with dimension higher than two, this correlation disappears, enabling the area-law, only in the strong-coupling limit, which corresponds to the leading order in $\beta\ll 1$. Indeed, this cannot occur in the $\beta\gg 1$ (weakly coupled) regime of $3+1$ dimensions electrodynamics, entailing the failure of the area-law and the establishment of a massless phase characterised by the potential \cite{Intriligator:1995er}
\begin{equation}
    V(R)=-\frac{{g^2(R)}}{R} + 2 V_0\,.
\end{equation}
This latter corresponds to a perimeter-law fall-off of the Wilson loop vev, i.e.
\begin{equation}
   \mathcal{W}_\alpha \sim \exp [-V_0 P_\alpha] \,,
\end{equation}
where the coupling constant $g(R)$ becomes a slow function of $R$ and tends to a constant in the Coulomb phase.\\

In a similar way, the magnetically disordered phase has been derived for non-abelian gauge theories while inspecting large Wilson loops, both by means of the Stokes theorem \cite{Arefeva:1979dp,Fishbane:1980eq,Diakonov:1989fc,Karp:1999vq,Hirayama:1999ar,Diakonov:2000kw,Kondo:2000pp,Kondo:1999tj} and by deploying finite-range correlators behaviours \cite{DiGiacomo:2000irz,Kuzmenko:2004hk}.

\subsection*{Center vortex mechanism for confinement}
\noindent 
A simple exposition of the center vortex mechanism of confinement was offered in
\cite{Engelhardt_1998}, and reviewed in \cite{Greensite_2017}. The argument is extremely simple and intuitive. We overview it in this section.\\

We start considering a planar rectangle of length $L$, which is pierced by $N$ center vortices. Embedded in the plane we consider a holonomy along a closed path $\alpha$, i.e.~a Wilson loop $W_\alpha$, enclosing an area $A$. The probability that $n$ among the $N$ vortices will be puncturing the area enclosed by the Wilson loop is provided by the generic factor of the Newton polynomial expansion, i.e.
\begin{equation}
P_N(n)= \left( \begin{array}{c}
     N  \\
     n 
\end{array} \right) \left( \frac{A}{L^2}  \right)^n \left( 1-\frac{A}{L^2}  \right)^{N-n}\,.
\end{equation}
The piercing of each vortex through the Wilson loop entails a contribution to the vev of the Wilson loop by a factor $(-1)$. Thus, the vev of a Wilson loop can be straightforwardly recovered by weighting the contributions provided by each piercing vortex with the probability distribution for $n$ vortices to puncture the area of the Wilson loop, namely
\begin{equation}
    \mathcal{W}_\alpha =\langle W_\alpha \rangle 
    = \sum \limits_{n=0}^N (-1)^n \, P_N(n) = \left(1-\frac{2A}{L^2} \right)^N\,.
\end{equation}
Introducing the vortex density $\rho=N/L^2$, which one may fix as constant in the infinite volume limit $L\rightarrow \infty $ and $N \rightarrow \infty$, it is immediate to recover the fall-off of the area-law:
\begin{equation}
   \mathcal{W}_\alpha =  \lim \limits_{N \rightarrow \infty} \left(1-\frac{2\rho A}{N} \right)^N= e^{-2 \rho A}\,.
\end{equation}
This mechanism will be at the onset of our formulation, based on out-of-equilibrium stochastic gauge-geometry flow and Topological Quantum Field Theory (TQFT). 

%%%%%%%%%%%%%%%%%%%%%%%%%%%%%%%%%%%%%%%%%%%%%%%%%%%%%%
\section{Theoretical framework}\label{sec:theory}
\noindent
%%%%%%%%%%%%%%%%%%%%%%%%%%%%%%%%%%%%%%%%%%%%%%%%%%%%%%
The theoretical framework our analysis is rooted in comprise topology and out-of-equilibrium physics. Topology is crucial for our discussions for the following reason. We associate to the hadronic background states generic (space-time) manifolds, on which the gauge and quark fields are supported, and the confined configurations are defined. We will argue below that precisely the stochastic (yet, classical) topology transformations instantiate the creation and annihilation of central vortices of the $SU(3)$ connection at the onset of the transitions among confined and deconfined phases in QCD.\\

On the other hand, creation and annihilation of center vortices, which are understood in terms of stochastic fluctuations in topology, require an out-of-equilibrium approach to the problem. Properties of relaxation toward the infrared hadronic background configurations that are characterized by confinement are derived by considering the out-of-equilibrium version of the Einstein--Yang--Mills equations. The effective action approach, with the calculation of loop-correction to the infrared Lagrangian of non-abelian gauge fields \cite{Savvidy:1977as}, is therefore substituted in this framework by a stochastic flow for the interacting physical systems that comprises the non-abelian gauge fields, the matter (quark) fields and the gravitational field, i.e.~geometry. The gravitational field, at the energy scales that are proper of the infrared physics and relevant to unveil confinement in hadronic states, is treated classically. Nonetheless, geometry is still crucial, since it drives the stochastic fluctuations of the underlining topology that enable to create/annihilate center vortices of the gluon fields.\\

Considering a (stochastic) geometric flow is therefore essential to define the topology fluctuations of the manifolds associated to hadronic background states. Stochasticity is a necessary requirement for the geometric flow, as it entails breakdown of diffeomorphisms in the infrared/large scale limit, the one at which the gravitational field is measured. Following the seminal paper by Parisi and Wu~\cite{Parisi:1980ys}, out-of-equilibrium physics is related to a breakdown of the infrared symmetry that is dynamical in the thermal (``stochastic'') time. The relevance of the Parisi-Wu approach for gravity was further addressed by Rumpf \cite{Rumpf:1985eh}, who extended the Parisi--Wu prescription for out-of-equilibrium stochastic quantization to the Einstein gravity. What is relevant for geometry in our approach is to describe the breakdown of the diffeomorphism symmetry (and the related changes in topology of the manifolds associated to the hadronic systems) of the gravitational field.\\

On the other hand, the size of these manifolds is still regulated by the dimensional transmutation scales of non-abelian gauge fields. The fermi scale that dictates the effective size of the hadronic system is not affected directly by the stochastic geometric flow.

%%%%%%%%%%%%%%%%%%%%%%%%%%%%%%%%%%%%%%%%
\subsection*{Stochastic geometric flow}
\noindent 
%%%%%%%%%%%%%%%%%%%%%%%%%%%%%%%%%%%%%%%%
The key idea is based on the fact that in the stochastic geometric (Ricci-like) flow the curvature fluctuations act as a multiplicative noise on the metric field,
\begin{equation}
        \frac{\mbox{d}}{\mbox{d}\tau} g_{\mu\nu} = -2 \imath R_{\mu\nu} + \xi_{\mu\nu} = -\imath \mathcal{G}_{\mu \nu \alpha \beta }\frac{\delta S_G}{\delta g_{\alpha\beta}} + \xi_{\mu\nu}  \,,
\end{equation} 
% \begin{equation}
%     \begin{split}
%         \frac{\mbox{d}}{\mbox{d}\tau} g_{\mu\nu} &= -2 \imath \left[R_{\mu\nu} - \frac{1}{2} g_{\mu\nu} R \right] + \xi_{\mu\nu} \\
%         &= -2\imath \mathcal{G}_{\mu \nu \alpha \beta }\frac{\delta S_G}{\delta g_{\alpha\beta}} + \xi_{\mu\nu}  \,,
%     \end{split}
% \end{equation} 
where $\xi_{\mu \nu}=\xi g_{\mu \nu}$ and $\mathcal{G}_{\mu \nu \alpha \beta }$ denotes the super-metric \cite{lulli2022stochasticquantizationgeneralrelativity}. The multiplicative noise $\xi_{\mu \nu}$ can be understood as a possible mechanism for multi-fractality, i.e.~the non-trivial scaling of the observables cumulants as a function of the scale. This is reflected in a fractal dimension that depends on the scale \cite{lulli2022stochasticquantizationgeneralrelativity}.\\ 

%Given that the tensorial character of the gravitational field is defined against general coordinate transformations, it is convenient to recall the transformation properties of the stochastic time $\tau$, which the reader can find in further details in \cite{lulli2022stochasticquantizationgeneralrelativity}. A possible choice is to interpret $\tau$ as the proper time. Its basic property is that it coincides with the coordinate time whenever the metric is almost Minkowskian, as it happens at spatial infinity for the Schwarzschild metric. This renders the equation covariant under general coordinate transformations, given that the proper time is a scalar. As in well-known considerations of turbulence, one can study the fluctuations of the metric components $\delta g_{\mu\nu}(\delta\tau) = g_{\mu\nu}(\tau_2) - g_{\mu\nu}(\tau_1)$ at a fixed proper time distance $\delta \tau = \tau_2 - \tau_1$, which sets a characteristic scale.\\

A necessary step for our analysis is to introduce a coupling to matter fields. This is achieved by considering the ``target'' Ricci tensor induced by the matter fields
\begin{equation}
    R^T_{\mu\nu} = \frac{8\pi G}{c^4}\left[T_{\mu\nu} - \frac{1}{2} g_{\mu\nu} T\right] \,,
\end{equation}
which enables to rewrite the stochastic geometric flow equation as follows:
\begin{equation} \label{TM}
        \frac{\mbox{d}}{\mbox{d}\tau} g_{\mu\nu} = - 2\imath [R_{\mu\nu} - R^T_{\mu\nu}] + \xi_{\mu\nu}\,.
\end{equation}
% \begin{equation} \label{TM}
%     \begin{split}
%         \frac{\mbox{d}}{\mbox{d}\tau} g_{\mu\nu} &= - 2\imath [R_{\mu\nu} - R^T_{\mu\nu}] + \xi_{\mu\nu}\\
%          &= - 2 \imath \left[R_{\mu\nu} - \frac{1}{2}g_{\mu\nu} R - \frac{8\pi G}{c^4} T_{\mu\nu} \right] + \xi_{\mu\nu} \,.
%     \end{split}
% \end{equation}
Here, we utilize the ansatz about the multiplicative nature of the gravitational stochastic noise, $\xi_{\mu\nu}\equiv \xi g_{\mu\nu} $, and relate it to the stochastic fluctuations of the trace of the Einstein equations at equilibrium. Instead, the square bracket in the second line of Eq.~\eqref{TM} still corresponds to the first variation of the action describing gravity and its coupling to matter.\\

The scaling dimension of the cumulants ($n$-points functions) could provide a possible mechanism for a dimensional reduction based on the intermittency of the curvature/matter fluctuations at the small scales \cite{lulli2022stochasticquantizationgeneralrelativity}. Knots can arise on average over large thermal time scales from configurations that might not be knotted at smaller scales, just because the effective (fractal) dimension would be different.

%%%%%%%%%%%%%%%%%%%%%%%%%%%%%%%%%%%%%%%%%%%%%%%%%%%%%%%%%%
\subsection*{Out-of-equilibrium Einstein--Yang--Mills}
\noindent 
%%%%%%%%%%%%%%%%%%%%%%%%%%%%%%%%%%%%%%%%%%%%%%%%%%%%%%%%%%
The Einstein--Yang--Mills (out-of-equilibrium) dynamics, which involves the mutual back-reaction of gauge fields and gravity, is the key to unveil the generation of center vortices. This finally provides the emergence of confinement through configurations supported on non-trivial topologies.

We first introduce the action for the gluon fields which are allowed to fluctuate out of equilibrium in the non-perturbative regime. The gluon fields then acquire a dependence on the thermal time, i.e.~$A^a_{\mu}(x,\tau)$, while $a$ denotes usual indices of the adjoint representation of the gauge symmetry group. On a curved space-time with Lorentzian signature $(-,+,+,+)$ and metric $g_{\mu \nu}(x)$, the Einstein--Yang--Mills action is expressed by
\begin{eqnarray}
\mathcal{S}_{\rm EYM}  = &&+ \frac{1}{2\kappa} \int d^4x \,\sqrt{-g}\,  R[g_{\mu\nu}] \nonumber  \\
&&- \frac{1}{4 g_s^2} \int d^4x \, \sqrt{-g} \, F^a_{\mu\nu} \, F^{a\, \mu\nu}\,.
\end{eqnarray}
For the associated Langevin equations written to describe the out-of-equilibrium dynamics of the system, we implement a multiplicative ansatz for the noise, along the lines of \cite{lulli2022stochasticquantizationgeneralrelativity}. The stochastic flow of the gauge field's components $A^a_\mu$ is introduced via the stochastic noise~$\xi^a_\mu(x,\tau) \equiv \xi'(x,\tau) A^a_\mu(x,\tau)$. Similarly, for the metric field components the stochastic noise term is denoted as~$\xi_{\mu \nu}(x,\tau)\equiv \xi(x,\tau) g_{\mu \nu}(x,\tau)$. Therefore, the equations of motion take the Langevin form
\begin{eqnarray}
\frac{\partial A^a_{\mu}(x,\tau)}{{\partial}\tau} = && \!\!\!\! \phantom{a} \imath \nabla^{\nu}F^a_{\nu\mu}(x,\tau) + \xi^a_\mu(x,\tau)\,, \label{F1}\\
\frac{\partial g_{\mu\nu}(x,\tau)}{{\partial}\tau} = &&  \!\!\!\! - \imath \frac{\sqrt{-g}}{2\kappa} \Big[ R_{\mu\nu}(x,\tau) - R^T_{\mu\nu}(x,\tau) \Big] \nonumber \\+ \xi_{\mu \nu}(x,\tau) \,. 
\label{F2}
\end{eqnarray}
Here, the gauge fields and their related strength fields are defined as usual by 
\begin{eqnarray}
\nabla^{\nu}F^a_{\nu\mu} && \!\!\!\!= \nabla_{\nu}F^{a \nu}{}_{\mu} \nonumber\\
&&\!\!\!\!=\partial_{\nu}F^{a \nu}_{\mu} + \Gamma^{\nu}_{\nu\rho}F^{a \rho}_{\mu} - \Gamma^{\rho}_{\nu\mu}F^{a \nu}_{\rho}\,,\\
F^a_{\mu\nu} && \!\!\!\!= \nabla_{\mu}A^a_{\nu} - \nabla_{\nu}A^a_{\mu} + f^{a}_{\ b c} A^b_{\mu} A^c_{\nu} 
 \nonumber\\
&& \!\!\!\! = \partial_{\mu}A^a_{\nu} - \partial_{\nu}A^a_{\mu} + f^{a}_{\ b c} A^b_{\mu} A^c_{\nu} \,,
\end{eqnarray}
with $f^{abc}$ structure constants of the gauge group, and the geometric quantities fulfill the standard definitions 
\begin{eqnarray}
&& \Gamma^{\mu}_{\nu\rho} = \frac{1}{2} g^{\mu\sigma} (\partial_{\nu}g_{\sigma\rho} + \partial_{\rho}g_{\nu\sigma} - \partial_{\sigma}g_{\nu\rho})\,,\\
&& R_{\mu\nu} = \partial_{\rho}\Gamma^{\rho}_{\mu\nu} - \partial_{\nu}\Gamma^{\rho}_{\mu\rho} + \Gamma^{\sigma}_{\mu\nu}\Gamma^{\rho}_{\sigma\rho} - \Gamma^{\sigma}_{\mu\rho}\Gamma^{\rho}_{\nu\sigma}\,,
\end{eqnarray}
with $R \equiv  g^{\mu\nu} R_{\mu\nu}$. Finally, the energy momentum tensor of the theory is defined by 
\begin{eqnarray}
&& T_{\mu\nu} = \frac{1}{4 g_s^2}F^a_{\mu\rho}F_{\nu}{}^{a \rho} - \frac{1}{16 g_s^2}F^a_{\rho\sigma}F^{ a \rho\sigma}g_{\mu\nu}\,,
\end{eqnarray}
the trace of which, $T \equiv  g^{\mu\nu} T_{\mu\nu}$, equates $R$ at equilibrium.\\

We argue that the out-of-equilibrium Einstein--Yang--Mills system in Eqs.~\eqref{F1}-\eqref{F2} describes stochastic topology fluctuations in the background manifold, which induce the appearance of center vortices. The appearance of the latter ones, being dual to Wilson loops, is specifically responsible for the asymptotic area-law scaling of Wilson-loop vevs. On the other hand, the appearance of center vortices, as induced by the stochastic gauge-geometry flow, can be interpreted as the emergence of closed chromo-magnetic Faraday lines dual to strings --- the chromo-electric fluxes between quarks. Both 't Hooft loops, responsible for the creation of center vortices, and Wilson loops evolve in the stochastic time $\tau$. The (stochastic) breakdown of the diffeomorphism symmetries entails the generation of topological defects, which can be in turn interpreted as vortices. On the other hand, the stochastic dynamics also induces singularities in the gauge transformations, which are also responsible for the creation of chromo-magnetic vortices as mentioned above. The two stochastic processes are related, and therefore we argue that also the stochastic noise amplitudes for the gravitational and gauge field must be related, consistently with a common origin due to chaos effects \cite{asselmeyermaluga2016smoothquantumgravityexotic}. It is also relevant to observe that chromo-electric and chromo-magnetic fluxes tangle through an Aharonov-Bohm (geometric) phase. The phase is due to the winding (in space-time) of the chromo-magnetic fluxes around the chromo-electric Wilson loops.

\subsection*{Singularities of the Ricci flow}

To be more specific, we provide below an
example of change of topology induced by the stochastic geometric flow. It is known that the Ricci flow can produce singularities when the scalar curvature is negative \cite{perelman2002entropy,kleiner2008notes}. This is realized by means of manifold surgery \cite{perelman2003ricci}. The idea of the Ricci flow with surgery is to remove the singularity fibers of a product of type $(-c,c)\times \mathbb S^3$ where the singularity is forming, and cap off $\{-c,c\}\times \mathbb S^3$. We therefore have (at finite time $T$) the formation of a neck-pinch singularity. We can then find a sequence of times $t_n$, where $t_n \longrightarrow T^-$ such that $\lim_n |{\rm Rm}| \longrightarrow +\infty$, where $|{\rm Rm}|$ is the largest absolute value of the sectional curvature. In the stochastic geometric flow, the expectation values of sectional curvatures diverge as well, and singularities can form. In fact, a standard application of \^Ito's lemma \cite{Ito1951} gives that the expectation value of a generic observable $X_\tau$  during an Ito process, as in the stochastic geometric flow \cite{lulli2022stochasticquantizationgeneralrelativity}, evolves over time according to the equality
\begin{eqnarray}
    \lim_{\Delta \tau\rightarrow 0} \frac{E_{\tau+\Delta\tau}(X_{\tau+\Delta\tau})-E_{\tau}(X_\tau)}{\Delta\tau} = \mu_\tau(X_\tau),
\end{eqnarray}
where $dX_\tau = \mu_\tau(X_\tau)d\tau + \sigma_\tau(X_\tau) dW_\tau$. In the stochastic geometric flow it means that the expectation values for the metric $g$ evolve as in the Ricci flow. Therefore, the unboundedness of $|{\rm Rm}|$ implies that the expectation value of $|{\rm Rm}|$ is unbounded as well during the stochastic geometric flow, which corresponds to the development of a singularity in finite time. 
We assume that at the formation of singularities a surgery cobordism removes the neck-pinch. However, we do not restrict ourselves to capping off with balls, which implies that the topology of the manifold changes. The simplest surgery is the attachment of two $1$-handles $D^1\times D^3$ to two pairs copies of $\mathbb S^3$, corresponding to collapsed fibers for the formation of two different singularities. This process changes the fundamental group of the base manifold, introducing nontrivial loops wrapping around the attached handles. This nontrivial elements of the fundamental group $\pi_1(M)$, where $M$ is the ground manifold, correspond via the standard surjection $\pi_1(M) \longrightarrow \frac{{\rm Hol}_A}{{\rm Hol^0}_A}$ for a connection $A$, where ${\rm Hol}_A$ is the holonomy group and ${\rm Hol}^{0}_A$ denotes a component of the identity. In general, there is a non-trivial class arising, showing that the topology changed. Thus, the surgery procedure creates nontrivial oscillations in the topology. The nontrivial homotopy classes of $\pi_1(M)$ correspond to Wilson loops that wrap around vortexes, whose creation is due to topological fluctuations induced by the creation of singularities in the stochastic geometric flow. In the next sections, we will delve in more details into topology-change processes. Specifically, we will adopt an approach that utilizes pairs composed by 1-handles and 2-handles.\\

Within our out-of-equilibrium (stochastic) gauge-geometric flow perspective, the winding of Wilson lines can also be described, in a stochastic quantization approach along the line of \cite{Parisi:1980ys}, as a topological braiding due to the $\theta$-term in QCD. This opens a pathway to a category theory interpretation of confinement (cobordism), as mainly due to topological effects and topology changes in the stochastic gauge-geometry flow. Consistent with this is our educated guess that the linking number of gluon field lines is at the onset of confinement, being instantiated by stochastic topology transformations of the underlying manifold. When two quarks are moved one away from another, some magnetic closed lines are broken, but then immediately recombine. This induces a breakdown and recombination of the chromo-electric Wilson lines, which are eventually braided and intertwined to the magnetic flux lines, due to the Aharonov-Bohm effect.
Within this framework, we propose that manifolds with the most simple topology (for instance, $\mathbb{R} \times S^3\cong \mathbb{R} \times S^3$) can be conjectured to give rise to ground-states such as the ones of mesons and nucleons, once their dual Wilson loop are linked, and correspondingly sourced, by the 't Hooft loops. Transitions to more complex topologies can give rise to multi-quark resonances and other excited meson states.\\

\section{Topological defects induced by the gauge flow}\label{topo-defects-gauge-flow}
\noindent
%%%%%%%%%%%%%
The realization of this program enforces us to follow a 
slightly different approach. We now investigate the 
spatial component of the spacetime, the 3-manifold, and its
variation of topology. To this purpose, we modify the
Langevin equations (\ref{F1}) and (\ref{F2}) to be
\begin{eqnarray}
{\frac{\partial A^a_{m}}{{\partial}\tau}}(x',\tau) \!=\! && \!\!\!\! \epsilon_{\ m}^{nl} F^a_{nl}(x',\tau) + \xi^a_m(x',\tau)\,, \label{F1_3D}\\
\!\!\frac{\partial g_{mn}}{{\partial}\tau} (x'\!,\!\tau) \!=\! \!\!&&  \!\!\!\! \frac{1}{2\kappa} \Big[ R_{mn}(x'\!,\!\tau) \!-\! \nabla_{\!m} \!\nabla_{\!n} f \Big] \!+ \xi_{mn}(x'\!,\!\tau)\!,  \, \ \ 
\label{F2_3D}
\end{eqnarray}
where $x'=(x^1,x^2,x^3)$ denotes the spatial coordinates, $\tau$ is the stochastic time, and the indices are running over  $l,m,n=1,2,3$. The epsilon tensor is only nonzero for $\epsilon_1^{\ 23}=1$ and cyclic permutation, being antisymmetric in the exchange of indices. 
The function $f$ is the generator of the diffeomorphism for the 3-manifold $\Sigma$. 
Both equations are the gradient flow for a corresponding functional. 
In Eq.~\eqref{F1_3D}, $F^a_{nl}$ introduces the Chern-Simons functional, while in Eq.~\eqref{F2_3D} $f$ introduces the entropy functional of Perelman \cite{perelman2002entropy}. We will discuss in what follows a 4D version of this functional. According to Ref.~\cite{Flo:88}, Eq.~(\ref{F1_3D}) is equivalent to the instanton equation. This can be shown by identifying the 0th component of the strength tensor with $F_{0\mu}=dA_\mu/dt$, where the derivative with respect to the time coordinate $t$ is expressed as a covariant derivative. We then obtain
\begin{equation} \label{impo}
    F_{\mu\nu}=\epsilon_{\mu\nu}^{\rho\kappa}F_{\rho\kappa}+
    \xi_{\mu\nu}\,,
\end{equation}
which is the stochastic version of the instanton equation, with the epsilon tensor extended to 4D, introducing $\epsilon_{01}$. 
Eq.~\eqref{impo} is important for several reasons. As shown by Seiberg \cite{Sei:88}, the instantonic solutions dominate the 
non-perturbative regime of a supersymmetric version of QCD. 
Furthermore, the theory is induced by a gradient flow of a Chern-Simons functional on the 3-manifold, which is known to induce vortices and its condensation. More importantly, the stochastic term can change the topological class of the theory,
as we proceed to show now.

In the following discussion, we switch for compactness of the notation to the language of differential forms. Then the stochastic instantonic equation is given by
\begin{equation}
    F=\star F +\xi\,.
\end{equation}
The minimum of the Yang-Mills functional
\[
\int F\wedge \star F
\]
is the topological invariant (second Chern class), 
\[
c_2=\frac{1}{8\pi^2}\int F\wedge F\,,
\]
i.e. the solution of the instanton previously reported.\\

We may now discuss the singularities of the gauge flow.
As an example, we start considering the case of  electromagnetism, i.e. a $U(1)$ gauge theory. The gauge potential $A$ with field strength $F=dA$ has to fulfill the equations
\[
dF=0\,,\qquad d*F=*j\,,
\]
the 1-form $j$ denoting the source. The equation $dF=d^2 A=0$ is a direct consequence of the relation $d^2 =0$. In presence of a monopole, the equation must be modified into
\begin{equation}
\label{zks}
dF=j_m\,,
\end{equation}
$j_m$ denoting the current of the magnetic charges.
Eq.~\eqref{zks} can be fulfilled by singular currents, according to the relation (in complex analysis)
\[
d\left( \frac{dz}{z} \right)=d^2(\ln(z))=2\pi \imath \delta(z) dz\wedge d\bar{z}\,.
\]
Nonetheless, this relation can be also fulfilled by a non-trivial complex line bundle (or a non-trivial $U(1)$ bundle) over the 2-sphere $S^2$ (or equivalently on $\mathbb{R}^3\setminus 0$).
Thus, the singularity can be understood as originating from a map between a trivial and a non-trivial complex line bundle. 
If we consider the stochastic gauge flow for a $U(1)$ gauge theory, then a singularity in this flow will produce a non-trivial bundle, or a monopole. \\

We may now extend this idea to the stochastic gauge flow of a $SU(3)$ gauge theory. Singularities of this flow are singular bundle maps producing non-trivial $SU(3)$ bundles that are classified by the instanton number. 
We start focusing on the definition
of a differential $df$ of a singular map $f$, using the work by
Harvey and Lawson in Ref.~\cite{HarLaw:93}, and first describe the general situation.

Let $F$ and $E$ be two vector bundles over $M$.
Furthermore, let $\alpha:E\rightarrow F$ be a bundle map admitting
singularities, i.e. a subset $\Sigma\subset M$ where the (local)
map $\alpha_{x}:E_{x}\rightarrow F_{x}$ is not injective. Let $D_{E}$
and $D_{F}$ be the connections on $E$ and $F$, respectively. The
main problem is now that the inverse of the map $\alpha$ not always
exists. Instead, we have to define a map $\beta:F\rightarrow E$. At
first we assume such a map to exist. Then we define the push forward
connection $\vec{D}^{\alpha}$ by \[
\vec{D}^{\alpha}=\alpha\circ D_{E}\circ\beta+D_{F}\circ(1-\alpha\beta)\,.\]
 On the two bundles, we may introduce metrics that allow to define the adjoint
$\alpha^{*}$ of $\alpha$ via the scalar product. Then, outside
the singular set $\Sigma$, we define \[
\beta=(\alpha^{*}\alpha)^{-1}\alpha^{*}\,.\]
 In general this procedure breaks down on the singular set $\Sigma$,
since $\beta$ becomes singular on $\Sigma$. To amend this situation, we choose
an approximation mode, i.e. a fixed smooth function $\chi:[0,\infty]\rightarrow[0,1]$, 
with $\chi'\leq0,\chi(0)=0$ and $\chi(\infty)=1$. We then define a smooth
approximation $\beta_{s}$ to $\beta$ by mean of the relation 
\[
\beta_{s}=\chi\left(\frac{\alpha^{*}\alpha}{s}\right)\beta%\,,\qquad\mbox{for $s>0$}
\,,\]
with $s$ some real positive parameter. The family of maps $\beta_{s}$ defines a family of connections
$\vec{D}_{s}^{\alpha}$ on $F$. As $s\rightarrow0$, the map $\beta_{s}$
converges to $\beta$ uniformly on any compact subset in $M-\Sigma$. The family
$\beta_{s}$ of maps converges for $s\rightarrow\infty$ pointwise
to a connection on $F$ for all points in $M$ --- see e.g. Ref.~ \cite{HarLaw:93}
for proofs.

Within this general context, we can define the
connection change by setting $E$ and $F$ to the trivial and non-trivial $SU(3)$ bundle, and $\alpha=df$ to represent the defect induced by the gauge flow. Then
the substitute for the inverse $df^{-1}$ is defined by the limit
$s\rightarrow\infty$ of $\beta_{s}$ constructed above. \\

We may now apply this theory to the singularities of the gauge flow. 
Let $E_0$ be a trivial $SU(3)$ bundle over the 4-sphere, regarded as a compactified $\mathbb{R}^4$. We can also consider the Minkowski space $\mathbb{R}^{3,1}$, with compactification $S^3\times S^1$. We may now consider the bundle $E_0$ over the spacetime $M$. According to the theory of Harvey and Lawson, every singularity of the gauge flow is related to a (singular) bundle map $a:E_0\to E_1$ from the trivial bundle $E_0$ to the non-trivial bundle $E_1$. The corresponding invariant is the second Chern class
\[
c_2(E_1)=\frac{1}{8\pi^2}\intop_M tr(F\wedge F)\,,\qquad c_2(E_0)=0
\]
for the curvature $F$ of the bundle $E_1$. 

According to this procedure, we may derive the  curvature
\begin{equation}
F=d\left( g_a^{-1} dg_a \right)
\label{singular-curvature}
\end{equation}
induced by the singular gauge transformation $g_a$ from the bundle map $a:E_0\to E_1$. The expression for the curvature represents the singular part of $d\left( g_a^{-1} dg_a \right)$, the other part $g_a^{-1} dg_a\wedge g_a^{-1} dg_a$ being neglected according to the Maurer-Cartan equation.\\

This formula has two fundamental consequences. Every singularity produces a non-trivial instanton --- i.e. a solution of the equation $F=\pm *F$. Then there is a simple relation between the holonomy along a closed loop $\gamma $ around the singularity generated by
\begin{equation}
\oint_{\gamma=\partial S} A=\intop_S dA=\intop_S d\left( g_a^{-1} dg_a \right)=\intop_S F\,,
\label{singular-holonomy}
\end{equation}
with $A=g_a^{-1} dg_a$. The integral over the curvature $F$, calculated according to (\ref{singular-curvature}), is generated by singularity, having integrated over a surface $S$, with boundary $\gamma$. The surface $S$ always exists, and is known as the Seifert surface. The loop $\gamma$ cannot be contracted, thus the integral over the surface does not vanish. This is still not enough to derive the scaling behaviour of the integral. Hence, we also require the stochastic geometry (Ricci) flow.\\

Summarizing, to each singularity we constructively associate an instanton. On the other hand, every solution $F$ also induces a unique solution $*F$. Consequently, for every color-electric loop (or Wilson loop), there must exist a color-magnetic loop (or t'Hooft loop), where both loops are linked.

\section{Out-of-equilibrium dynamics and Renormalization Group}\label{sec:out-of-eq}
\noindent 
%%%%%%%%%%%%%%%%%%%%%%%%%%%%%%%%%%%%%%%%%%%%%%%%%%%%%%%%%%%%%%%%%%%%
The connection between standard RG flow techniques deployed in QFT to investigate the running of gauge and diffeo-invariant observables, and the evolution of the Einstein--Yang--Mills system in a stochastic time deserves a separate analysis that is currently ongoing. Nonetheless, a few interesting features can be immediately derived that concern the appearance in the out-of-equilibrium phase of torsional degrees of freedom, possibly with relevant phenomenological consequences. \\ 

What one can see in the case of gauge theories, already in the original paper of Parisi--Wu \cite{Parisi:1980ys}, is that the divergences arising from the singular gauge transformation seem to be translated directly to divergences in the stochastic-time dependence. These divergences are then similar to the secular terms in the multi-scale expansions of differential equations. If we choose the ``extended'' connection in the projective form
\begin{equation}
\bar{\Gamma}^\alpha_{\beta\gamma} = \Gamma^\alpha_{\beta\gamma} + \delta^\alpha_\gamma U_\beta \,,
\end{equation}
we immediately derive a conformal transformation of the metric tensor that is sourced by the gradient flow:
\begin{equation}
\frac{\text{d}}{\text{d}\tau}g_{\mu\nu}\left(\tau\right)= - 2 n^\alpha (\tau) U_\alpha (\tau) g_{\mu\nu}(\tau) = -2\lambda\left(\tau\right)g_{\mu\nu}\left(\tau\right)\,.
\end{equation}
This choice of the extended connection is indeed consistent with a rescaling of the metric as a function of the thermal time and the space-time coordinates, with a scaling factor provided by $\lambda(\tau) = n^\mu(\tau) U_\mu(\tau)$, which is a local variable.\\

We may then ask ourselves whether there exists a relation between the scale factor $\lambda$ of the Ricci flow and the scale factor of the RG flow. For sure, the value of the scale factor in QFT is not a field itself, i.e.~it does not depend on the coordinates. Instead, it would coincide in the case of the Minkowski metric, which is homogeneous and isotropic, with $\lambda$ turning out to depend on $\tau$ only. In a sense we would be ``deparametrizing'' the scale, in a similar way as one deparametrizes time in General Relativity.\\

Another relevant point concerns how to recover the (deparametrized) RG flow of QFT in the low-curvature limit. This was addressed in Ref.~\cite{Lulli:2023fcl}, in light of its relevance for the foundational issue of the collapse of the wave function. This latter was also investigated through the lenses of the stochastic Ricci flow in Ref.~\cite{Lulli:2023fcl}, and recognised to be induced by the back-reaction with the geometry. \\

Finally, we mention results that will appear in \cite{Lulli_2024B}, concerning the role of the torsional degrees of freedom in stochastic geometric flows. Specifically, it is possible to show that perturbations of the metric that are out-of-equilibrium introduce a projective torsional term to the connection. Out-of-equilibrium, torsion is dynamical, and therefore the degrees of freedom that are related to it will have to evolve according to the Langevin equation. Asymptotically, the relaxation toward equilibrium that enables to obtain General Relativity is achieved through a corresponding evaporation of the torsional degrees of freedom. Thus, torsional degrees of freedom can be interpreted as a byproduct of the stochasticity of the underlined geometry, and are unfrozen in the out-of-equilibrium phase. Consistently with this picture, the extended connection induces a scale transformation on the reference metric tensor generated by its proper-time derivative.

\section{Topology changes from the stochastic geometry flow}\label{sec:wilson}
\noindent
In this section we will analyse the stochastic geometry (Ricci) flow and its implications for the topology of the Wilson loop. 
In section \ref{topo-defects-gauge-flow} we discussed the appearance of Wilson loops and t'Hooft loops due to the singularities of the gauge flow.
The gauge flow is coupled to the geometry flow. Then, the stochastic geometry flow induces fluctuations of the Wilson loops by the fluctuating metric of the embedding space of the loop.
We may now discuss the possible fluctuations of the Wilson loops. At first, we clarify the notation. The Wilson loop is an unknotted solid torus: we use the word loop or unknot to express the unknotness. 
The name ``Wilson knot'' will be used in the following to indicate a knotted Wilson loop represented by a knotted solid torus. In a dynamical picture, the Wilson knot must fluctuate due to the stochastic geometry flow. For small fluctuations, the Wilson knot does not change and it is topologically stable. However, for large fluctuations, two parts of the closed curve come close to each other, so that the two parts meet and rearrange in a different way. Then a trivial Wilson knot (i.e. a Wilson loop) changed to a knotted Wilson loop (i.e. a Wilson knot). 
This operation is well-known in knot theory: one obtains a new knot that is concordant to the previous knot. There is a result that every knot is changed by the concordance relation to a hyperbolic knot, i.e.~a knot where the knot complement defined below admits a hyperbolic geometry \cite{Myers1983}.\\

We may now consider the close neighborhood around a Wilson loop, i.e.~a solid torus embedded in $M$, with $\gamma$ being its meridian. We can then consider the knot complement $C(\gamma) = S^3 - \gamma\times D^2$. This represents the geometry around the knot, i.e. how the knot has an impact on the geometry of the space. 

This brings to the following scenario. At low energies one has small fluctuations. The Wilson loop is determined by the unknotted, closed curve (i.e. the unknot or loop), and the loop is supported by the length of the curve. In contrast, large fluctuations induce a change of the Wilson knot, so that it is necessary to consider the corresponding change of the knot induced by the cobordism between the two knots. Then the Wilson knot is determined by the surface, i.e.~the cobordism. This behavior leads to the confinement \cite{Wilson:1974sk}. The size of this surface is constrained by the fact that every knot is concordant to a hyperbolic knot, thus the size of surfaces in hyperbolic manifolds (the knot complement) is constrained by Mostow rigidity. This is the outline of what we specify below.\\

We first provide a possible interpretation of the stochastic noise as a stochastic diffeomorphism. At this purpose, we consider a function $f:M\to \mathbb{R}$ on the 3-manifold and the geometry flow defined by
\begin{equation}
    \frac{\mbox{d}}{\mbox{d}\tau} g_{\mu\nu} = - 2R_{\mu\nu}-2\partial_\mu \partial_\nu f \,,
\end{equation}
where $f$ is considered as the generator of diffeomorphisms. This equation can be written as the gradient flow of the functional
\begin{equation}
    S=\int_M e^f(R+|\nabla f|^2) \ d{\rm vol}(M) \,,
\end{equation}
where $f$ can then be interpreted as dilaton. Usually, a stochastic term is only a continuous function. Nonetheless, every homeomorphism of the 3-manifold can be approximated with arbitrary accuracy by a diffeomorphism (topological and smooth 3-manifolds are equal). Alternatively, one can also consider a $C^2$ function, in such a manner that $\partial_\mu\partial_\nu f$ is only continuous.\\

As a second point, we have to consider the (potential) convergence behavior of the geometry flow. The behavior depends strongly on the topology of the 3-manifold and its underlying geometric structure. In Appendix~\ref{sec:topology} we describe the eight possible geometries on 3-manifold and the decomposition of the 3-manifold into geometric pieces. We only state here the general behavior: every spherical geometry has a finite convergent time in the geometry flow --- except for the hyperbolic geometry, the other geometries lead to a pinching of the underlying 3-manifold. Therefore, only the geometry flow for the hyperbolic geometry can be written as a stochastic equation away from the fixed point.

We may now go a step further. We discussed in section \ref{topo-defects-gauge-flow} the pairwise appearance of a Wilson knot and a t'Hooft loop. This configuration has a simple interpretation. Let us focus on the trivial case, in which one Wilson loop and one t'Hooft loop produce the 3-sphere. This decomposition is known as Heegard decomposition of the 3-sphere $S^3=(D^2\times S^1)\cup (S^1\times D^2)$. One can understand this decomposition by a simple argument. The 4-disk $D^4$ has the 3-sphere as boundary $\partial D^4=S^3$. We then use the relation $D^4=D^2\times D^2$ and build the boundary $S^3=\partial D^4=\partial (D^2\times D^2)$,  to derive $\partial (D^2\times D^2)=(\partial D^2 \times D^2)\cup (D^2\times \partial D^2)$, leading to the final result. 

For the case in which a loop (or unknot) is changed to a knot, then the decomposition can be also applied. Nonetheless, one obtains a homology 3-sphere, i.e. a 3-manifold with the same homology than a 3-sphere. This procedure is known as $\pm 1$ Dehn surgery along a knot. Then the change of the loop to a knot can be also expressed as a cobordism from the 3-sphere to a homology 3-sphere. 

We are now able to determine the scaling behavior of the Wilson functional. At this purpose, we use the geometry of the knot complement. As previously discussed, most knot complements admit a hyperbolic geometry. Then the surface that is bounded by this knot (the Seifert surface) has a fixed area (determined by Mostow rigidity). 

Furthermore, the knot complement has a geometry of constant curvature, which affects also the curvature of the Wilson knot. We then derive the desired behavior
\[
\intop_S F=\intop_S d(g_a^{-1} dg_a)\approx  {\rm Area}(S)
\]
for large fluctuations or large curvatures. For small fluctuations one cannot find  such a surface, and the integral scales only w.r.t. length of the loop.

\section{Scaling formula for the topology change}
\noindent
In this section, we discuss the stochastic fluctuations of the geometry flow, as expressed in terms of the fluctuations of the Wilson loop that lead to the topology change. As we clarified earlier, the stochastic geometry flow is connected with Einstein equations. Thus the energy scales that will naturally play a role will be the Planck scale, because of the appearance of the Newton (coupling) gravitational constant in the equation of motion at equilibrium, and the square root of the cosmological constant, to which the stochastic noise is linked \cite{lulli2022stochasticquantizationgeneralrelativity}. One should then require an interplay between these two energy scales, in order to produce a topology change in the system. Henceforth, we will focus on the direct consequences of the topology change as induced by the stochastic fluctuations of the 
Wilson loop. \\

The stochastic geometry flow leads to fluctuations of the Wilson loop. We allow for the topology change by using a cobordism, i.e. a 2-manifold connecting the unknotted Wilson loop into a knotted Wilson loop. This change is denoted as (ribbon) concordance. We may now ask ourselves what is the simplest topology change. Here, we need a 4-dimensional perspective to obtain an answer. We previously described the configuration in which a Wilson loop is linked to a t'Hooft loop as a 3-sphere. A change from a unknotted Wilson loop to a knotted Wilson loop is equivalent to a change from the 3-sphere to a homology 3-sphere. This change can be also described by a cobordism, known as homology cobordism. In the simplest case, one needs a pair of one 1-handle and one 2-handle to induce the simplest topology change. We now have to face the problem how to derive a scale for the homology 3-sphere from the scale of the 3-sphere. We first focus on gaining a conceptual understanding of this question, before proceeding to the calculation of the concrete value. The approach was partly developed in \cite{AsselmeyerKrol2018a}, where it was used to calculate the cosmological constant. Within the mechanism we discuss here, the cosmological constant is related to the stochastic noise.\\

The change of the topology from the 3-sphere $S^3$ to some homology 3-sphere $\Sigma$ is described by a smooth 4-manifold $W$ with boundary $\partial W=S^3 \sqcup \Sigma$, the so-called (homology) cobordism. Customarily, this change is described in the reversed way, i.e. the 1-handle and the 2-handle have to cancel each other. This process needs an embedded disk, which is complicated to be constructed in 4D. Instead, we need an infinite tower $CH$ of disks arranged along an infinite tree $T_{CH}$, usually called a Casson handle $CH$. The infinite tree representing this Casson handle must be embedded in a compact subset, which is only possible by using a subset with hyperbolic metric. For the embedding of the tree, it is enough to use a 2D model, i.e. the hyperbolic space $\mathbb{H}^{2}$. There are many isometric models of $\mathbb{H}^{2}$.
We will use later the Poincar\'e disk model, but for our current purpose it suffices to deploy the half-plane model, selecting the hyperbolic metric 
\begin{equation}
ds^{2}=\frac{dx^{2}+dy^{2}}{y^{2}}\label{eq:hyp-half-plane}
\end{equation}
in order to simplify the calculations. It is anyway possible to switch between the two isometric models, namely the Poincar\'e disk model and the half-space model. \\

The infinite tree $T_{CH}$ is embedded along the $y-$axis, so we may set $dx=0$. The tree $T_{CH}$, as the representative for the Casson handle $CH$, can be seen as a metric space instead as a simplicial tree. While focusing on a simplicial tree, one is only interested in the structure given by the number of levels and branches. The tree $T_{CH}$, as a metric space (the so-called $\mathbb{R}-$tree), has the property that any two points are joined by a unique arc isometric to an interval in $\mathbb{R}$. Then the embedding of $T_{CH}$ is provided by the identification of the coordinate $y$ with the coordinate of the tree $a_{T}$, representing the distance from the root. This coordinate is a real number, and we can build the new distance function, after the embedding, as 
\[
ds_{T}^{2}=\frac{da_{T}^{2}}{a_{T}^{2}}\,.
\]
As we have previously discussed, the tree $T_{CH}$ grows with respect to a time parameter, so that we need to introduce an independent ``time scale'' $t$. From the physics point of view, this time scale describes the partition of the tree into slices. Then a natural choice seems to be to set 
\[
ds_{T}^{2}\sim dt^{2}\,.
\]
Here the time scale is related to the hyperbolic distance 
\begin{equation}
\frac{da_{T}^{2}}{a_{T}^{2}}=\frac{1}{L^{2}}dt^{2}\label{eq:generalized-Friedman-equation}
\end{equation}
via the scale $L$ of the hyperbolic structure for the 3-manifold $Y_{n}$ (see below) with $n=[a_{T}]$. However this equality is only a heuristic argument. The 3-manifolds $Y_n$ are intermediate steps in the topology change processes induced by the corresponding level in the Casson handle, constrained by the values $Y_0=S^3$ and $Y_\infty=\Sigma$. 
A rigorous mathematical argumentation for the embedding of the Casson handle can be found in the Appendix~\ref{Casson}.

In the Casson handle $CH$, we have an infinite tower $CH=\bigcup_n T_n$ of disks (with self-intersections) arranged in levels $T_n$. For every level $T_n$, one has a scaling $L$. We then choose the time $t$ to be equal to the level $n=[t]$ for the integer part --- remember the tree is a metric tree. Therefore, we are forced to make the identification 
\[
Re(\xi)=\frac{t}{L}
\]
within the quadratic differential, defining the tree $T_{CH}$ --- see the Appendix~\ref{Casson}. Then the length in the embedded tree is given by $d(Re(\xi$)), the  measure of the vertical foliation. Now, the growth $ds_{T}^{2}$ of the tree with respect to the hyperbolic structure is provided by the measure $d(Re(\xi))^{2}$ of the vertical foliation, or 
\[
ds_{T}^{2}=\frac{da_{T}^{2}}{a_{T}^{2}}=d\left(\frac{t}{L}\right)^{2}\,,
\]
which is in agreement with previous heuristic, dimensional argument. This equation can
be formally integrated yielding the expression 
\begin{equation}
a_{T}(t,L)=a_{0}\cdot\exp\left(\frac{t}{L}\right)\label{eq:integration-general-Friedmann}\,.
\end{equation}
We shall now determine the ratio $t/L$.\\ 

How can we interpret this tree? The tree represents the process to change the Wilson loop as encoded into the topology change $S^3\to \Sigma$. This process is driven by the stochastic geometry flow. The possible changes of the Wilson loop, as driven by the noise in the stochastic geometry flow, can be in  principle arranged into a tree. There occur to be many small changes, and only a few larger changes. We hence interpret the tree in the following sense: the higher levels represent small changes, whilst the lower levels represent the larger changes. From the formal point  of view, we introduce intermediate steps between $S^3$ and $\Sigma$, denoted by $Y_n$. To every $Y_n$ we assign the length scale $L$. Then $Y_{\infty}$ must be identified with $\Sigma=Y_\infty$. We then remark that $Y_n$ are hyperbolic 3-manifolds with a fixed length scale called Mostow-Prasad rigidity --- the volume of a hyperbolic 3-manifold is a topological invariant. Then we define
\[
|^{3}R|=\frac{1}{L^{2}}
\]
to be the absolute value of the scalar curvature of $Y_{n}$ with
respect to the scale $L$. By a simple integration with respect to the metric $h$ of the 3-manifold $Y_{n}$, we obtain 
\[
\frac{1}{L^{2}}=\frac{\intop_{Y_{n}}\,|^{3}R|\sqrt{h}d^{3}x}{\intop_{Y_{n}}\sqrt{h}d^{3}x}\,,
\]
expressing the constant scalar curvature. In the following, we will relate the expression
\[
\intop_{\Sigma}{\rm Tr}\left(A\wedge dA+\frac{2}{3}A\wedge A\wedge A\right)=8\pi^{2}CS(\Sigma)\,,
\]
known as Chern-Simons invariant of a 3-manifold $\Sigma$, to the previous curvature integral. Using ideas of Witten \cite{Wit:89.2,Wit:89.3,Wit:91.2}, we will interpret
the connection $A$ as an $ISO(2,1)$ connection. Note that $ISO(2,1)$ is the Lorentz group $SO(3,1)$ by In{\"o}n{\"u}-Wigner contraction, or the
isometry group of the hyperbolic geometry. At this purpose, we choose
\begin{equation}
A_{i}=\frac{1}{\ell}e_{i}^{a}P_{a}+\omega_{i}^{a}J_{a}\,,\label{eq:Cartan-connection}
\end{equation}
with $\ell$ a length scale and $A=A_{i}dx^{i}$ a 1-form valued in the Lie algebra $ISO(2,1)$. Thus the generators $P_{a}$, $J_{a}$ fulfill the commutation relations 
\[
[J_{a},J_{b}]=\epsilon_{abc}J^{c}\,,\qquad[P_{a},P_{b}]=0\,,\qquad[J_{a},P_{b}]=\epsilon_{abc}P^{c}\,,
\]
with pairings $\langle J_{a},P_{b}\rangle={\rm Tr}(J_{a}P_{a})=\delta_{ab}$,
$\langle J_{a},J_{b}\rangle=0=\langle P_{a},P_{b}\rangle$. We then
obtain for the curvature 
\begin{eqnarray*}
F_{ij} & = & \frac{1}{\ell}P_{a}\left(\partial_{i}e_{j}^{a}-\partial_{j}e_{i}^{a}+\epsilon^{abc}
(\omega_{ib}e_{jc}+e_{ib}\omega_{jc})\right)+\\
 &  & J_{a}\left(\partial_{i}\omega_{j}^{a}-\partial_{j}\omega_{i}^{a}+\epsilon^{abc}\omega_{ib}\omega_{jc}\right)\,.
\end{eqnarray*}
In what follows, we move from the expression $A\wedge F$, and use the pairing $\langle\,,\,\rangle={\rm Tr}(\,)$ following the MacDowell-Mansouri approach ---  see \cite{MacDMan:1977}. Then we obtain 
\[
{\rm Tr}(A\wedge F)=\frac{1}{\ell}e\wedge(d\omega+\omega\wedge\omega)+\frac{1}{\ell}\omega\wedge(de+\omega\wedge e)\,,
\]
where the second term provides $\omega\wedge T$, with $T$ torsion. The first term encodes the structure $e\wedge R$, with curvature 2-form $R$, which reproduces the scalar curvature $^{3}R$ multiplied by the volume form (in the first order formalism) within the Einstein-Hilbert action. Therefore, for vanishing torsion $T=0$, we obtain 
\[
\intop_{\Sigma}{\rm Tr}(A\wedge F)=\frac{1}{\ell}\intop_{\Sigma}\,^{3}R\sqrt{h}d^{3}x\,.
\]
By performing an auxiliary scaling $\omega\to\frac{3}{2}\omega$, we obtain the
new connection 
\[
\frac{2}{3}A_{i}=\frac{2}{3\ell}e_{i}^{a}P_{a}+\omega_{i}^{a}J_{a}\,,
\]
and then finally the relation 
\[
{\rm Tr}\left(A\wedge(dA+\frac{2}{3}A\wedge A)\right)=\frac{3}{2\ell}e\, \wedge(d\omega+\omega\wedge\omega)\,,
\]
or 
\begin{equation}
8\pi^{2}\cdot\ell\cdot CS(\Sigma)=\frac{3}{2}\intop_{\Sigma}\,^{3}R\sqrt{h}d^{3}x\label{eq:CS-to-EH}\,.
\end{equation}

In (\ref{eq:Cartan-connection}) we have been forced to introduce the length scale $\ell$ for the translation generators $P_{a}$. This rescaling can now be parametrized by a coordinate
$t$. Therefore, we redefine the expression $8\pi^{2}\ell$ in (\ref{eq:CS-to-EH})
as 
\begin{equation}
t\cdot CS(\Sigma_{2})=\frac{3}{2}\intop_{\Sigma_{2}}\,^{3}R\sqrt{h}d^{3}x\label{eq:scalar-curvature-CS-inv}\,,
\end{equation}
where the extra factor $8\pi^{2}$ (which equals $4\cdot vol(S^{3})$) is the normalization of the curvature integral. This normalization of the curvature changes its absolute value into
\begin{equation}
|^{3}R_{\rm ren}|=\frac{1}{8\pi^{2}L^{2}}\label{eq:renormalized-curvature}\,.
\end{equation}
This naturally suggests a choice of the scale factor from its relation with the volume $L=\sqrt[3]{{\rm vol}(Y_{\infty})/(8\pi^{2})}$, where again $Y_{\infty}=\Sigma$. We finally obtain the formal relation 
\begin{equation}
\intop_{Y_{\infty}}\,|^{3}R_{\rm ren}|\sqrt{h}\,d^{3}x=\intop_{Y_{\infty}}\frac{1}{8\pi^{2}L^{2}}\sqrt{h}\,d^{3}x=L^{3}\cdot\frac{1}{L^{2}}=L\label{eq:CS-integral-relation}
\,,
\end{equation}
by using 
\[
L^{3}=\frac{{\rm vol}(Y_{\infty})}{8\pi^{2}}=\frac{1}{8\pi^{2}}\intop_{Y_{\infty}}\sqrt{h}\,d^{3}x\,,
\]
which is in agreement with the previous normalization. We may now use the absolute value of the curvature $|^{3}R|$ and of the Chern-Simons
invariant $|CS(\Sigma)|$. By (\ref{eq:scalar-curvature-CS-inv})
and (\ref{eq:CS-integral-relation}), and using 
\[
\frac{t}{L}=
\frac{3}{2\cdot CS(\Sigma)}\,, 
\]
and  Eq.~(\ref{eq:integration-general-Friedmann}), we derive 
the following exponential behavior 
\[
a(t)=a_{0}\cdot e^{t/L}=
a_{0}\cdot\exp\left(\frac{3}{2\cdot CS(\Sigma)}\right)\,.
\]
We may discuss the topology change $S^3\to \Sigma$. The change by adding a single pair of 1-/2-handle produces in the simplest case the so-called Mazur manifold with boundary the Brieskorn sphere $\Sigma=\Sigma(2,5,7)$. 
Then we will calculate the change in the scale by the above formula. As previously noted, we assume a Planck-size $(L_{P})$ --- natural (low-energy) scale for the geometry side --- for the 3-sphere $S^{3}$. This  changes to the scale $a$ of
$\Sigma(2,5,7)$ according to
\[
a=L_{P}\cdot\exp\left(\frac{3}{2\cdot CS(\Sigma(2,5,7))}\right)\,,
\]
with the Chern-Simons invariant
\[
CS(\Sigma(2,5,7))=\frac{9}{4\cdot(2\cdot5\cdot7)}=\frac{9}{280}\,.
\]
This provides a tremendous result: the energy scale, which is order of the Planck scale, i.e. $10^{-34}$ m, changes to value $10^{-15}$ m. In particular, we will obtain 
\[
a_{1}=L_{P}\cdot\exp\left(\frac{140}{3}\right)\approx7.5\cdot10^{-15}\,{\rm m}.
\]
This can be finally related to an energy scale. The estimate of its related Compton length provides the value of the relevant energy scale entering the topology changes processes that we speculate are at the origin of confinement, namely 165 MeV. This is indeed comparable to the energy scale
of QCD, which lies between 217 MeV and 350 MeV --- see e.g. Refs.~\cite{EnergyScaleQCD:2004,EnergyScaleQCD:2016}.

%%%%%%%%%%%%%%%%%%%%%%%%%%%%%%%%%%%%%%%%%%%%%%%%%%%%%%%%%%%%%%%%%%%%
\section{Proposal for QCD confinement}\label{sec:proposal}
\noindent 
%%%%%%%%%%%%%%%%%%%%%%%%%%%%%%%%%%%%%%%%%%%%%%%%%%%%%%%%%%%%%%%%%%%%
Geometric fluctuations due to the stochastic geometric flow induce changes in the field lines, following the pattern of knot concordance described in Section~\ref{sec:wilson}. Within this framework, the expectation of the knotted and linked Wilson loops is computed by considering the vev of the product of two Wilson loops as 
\begin{eqnarray}\label{eqn:wilson_link}
    \langle {W}_{\gamma_1}(A) {W}_{\gamma_2}(A) \rangle %=&& \frac{1}{Z}\int \mathcal{D} A\, {W}_{\gamma_1}(A) \, {W}_{\gamma_2}(A)  \\
    =&& \frac{1}{Z}\int \mathcal{D} A \, {\rm Tr}\left[\mathcal{P} e^{\imath \oint_{\gamma_1}A_\mu dx^\mu}\right] \times \nonumber \\
    && {\rm Tr}\left[\mathcal P e^{\imath\oint_{\gamma_2}A_\mu dx^\mu}\right]
    e^{\imath S_{\rm YM}[A]}\,,   %\nonumber
\end{eqnarray}
where $S_{\rm YM}$ denotes the Yang--Mills action, and the normalization constant $Z$ is the partition function. Delving into the analogy with the Chern-Simons case studied in Ref.~\cite{cotta1990quantum}, we expect $\langle {W}_{\gamma_1}(A){W}_{\gamma_2}(A) \rangle$ to depend on the linking of $\gamma_1$ and $\gamma_2$. Within the Chern-Simons framework, it is known that switching the crossings produces certain skein relations ---  see e.g.~Ref.~\cite{cotta1990quantum}. This explains why the vev of Wilson loops provides link invariants, such as the Jones polynomial.\\

In order to support this conjecture, we show how to estimate the path integral in Eq.~\eqref{eqn:wilson_link}. At this purpose, we move from Eq.~(\ref{singular-holonomy}), and use the relation in Eq.~\eqref{singular-curvature}. Then we recast the Wilson functional as
\[
{W}_{\gamma_1}(A)= {\rm Tr}\left[\mathcal{P} e^{\imath\oint_{\gamma_1}A_\mu dx^\mu}\right]={\rm Tr}\left[\mathcal{P} e^{\imath\intop_{\Gamma_1}F_{\mu\nu} dx^\mu \wedge dx^\nu}\right]\,,
\]
with the surface $\Gamma_1$ such that $\partial\Gamma_1=\gamma_1$.
Schematically, we obtain
\[
{W}_{\gamma_1}(A)= {\rm Tr}\left[\mathcal{P} e^{\imath \intop_{\Gamma_1}F\, d{\rm vol}(\Gamma_1)}\right]\,,
\]
with $F$ denoting the curvature of the connection $A$. The Yang-Mill action consists of the term
\[
g^{\mu\rho}g^{\nu\lambda}F_{\mu\nu}F_{\rho\lambda}\,,
\]
which we write schematically as $g^2F^2$. Thus we obtain the integral
\[
S_{YM}=\intop_M g^2 F^2 d{\rm vol}(M)
\]
over the spacetime $M$. To proceed with the calculation, we have to introduce the support function ${\rm supp}(\Gamma_1)$, with
\[
\intop_M F\, {\rm supp}(\Gamma_1)d{\rm vol}(M) = \intop_{\Gamma_1} F d{\rm vol}(\Gamma_1)\,.
\]
The following relations are straightforwardly recovered:
\begin{eqnarray*}
\intop_M {\rm supp}(\Gamma_1)\, d{\rm vol}(M)&=&\intop_{\Gamma_1}\, d{\rm vol}(\Gamma_1)={\rm Area}(\Gamma_1) \,,\\
({\rm supp}(\Gamma_1))^2&=&{\rm supp}(\Gamma_1)\,.
\end{eqnarray*}
We may now recast the Wilson functionals and the Yang-Mills action in one integral over $M$, with kernel
%\[
$F^2 g^2 + ({\rm supp}(\Gamma_1)+{\rm supp}(\Gamma_2))F\,.
%\]
$
This latter can be further recast according to
\begin{equation}
\left(\!Fg\!+\!\frac{({\rm supp}(\Gamma_1)\!+\!{\rm supp}(\Gamma_2))}{2g} \right)^2\!\!-\frac{\left({\rm supp}(\Gamma_1)+{\rm supp}(\Gamma_2)\right)^2}{(4g^2)}\,.
\end{equation}
From first expression, which is quadratic in the field strength, we can directly proceed to calculate the path integral. While from the second expression we obtain (up to constants)
\begin{eqnarray*}
\intop_M \left({\rm supp}(\Gamma_1)+{\rm supp}(\Gamma_2)\right)^2 d{\rm vol}(M)= \\
={\rm Area}(\Gamma_1)+{\rm Area}(\Gamma_2)+2\intop_M {\rm supp}(\Gamma_1){\rm supp}(\Gamma_2) d{\rm vol}(M)\,.
\end{eqnarray*}
The latter integral can be interpreted as the intersection $\Gamma_1\pitchfork\Gamma_2$ between the two surfaces $\Gamma_1$ and $\Gamma_2$. This is equivalent to the linking number $lk(\gamma_1,\gamma_2)$ between the knots $\gamma_1$ and $\gamma_2$ weighted by volume $vol(\Gamma_1\pitchfork\Gamma_2)$ of the intersection, i.e.
\[
\intop_M {\rm supp}(\Gamma_1){\rm supp}(\Gamma_2) d{\rm vol}(M)=lk_{vol}(\gamma_1,\gamma_2)\cdot vol(\Gamma_1\pitchfork\Gamma_2)\,,
\]
where ``$lk_{vol}(\gamma_1,\gamma_2)$'' denotes the weigthted linking number
\[
lk_{vol}(\gamma_1,\gamma_2)= lk(\gamma_1,\gamma_2)\cdot vol(\Gamma_1\pitchfork\Gamma_2)
\]i.e. linking number $lk(\gamma_1,\gamma_2)$ between the knots $\gamma_1$ and $\gamma_2$ multiplied with the volume $vol(\Gamma_1\pitchfork\Gamma_2)$. Consistently with these relations, we derive up to constants the behaviour  
\begin{equation}
    \langle {W}_{\gamma_1}(A) {W}_{\gamma_2}(A) \rangle
    \sim e^{\imath\, {\rm Area}(\Gamma_1)+\imath\,{\rm Area}(\Gamma_2)}\, 
    e^{2\imath\, lk_{vol}(\gamma_1,\gamma_2)}\,,
\end{equation}
which is our desired result. %The details of this calculation will be presented in a forthcoming work.

%\TAM{
%HAS TO BE REWRITTEN:  In the case of Yang--Mills, we do not expect the result to produce link invariants, but we conjecture that the linking results in interactions that influence confinement. Since linking is a property induced by stochastic fluctuations, as suggested in this article following \AM{the same strategy developed for the stochastic RF, adapted to the stochastic gauge-geometry flow}, it follows that confinement properties are induced by geometric properties. While we defer a detailed study of Eq.~\eqref{eqn:wilson_link} and its relation to confinement to subsequent studies, we emphasise that one can compute $\langle W_{\gamma_1}(A)W_{\gamma_2}(A) \rangle$ also in the case where $\gamma_2$ is obtained from $\gamma_1$ by a small shift, as in the case of the self-linking number. This corresponds to the presence of self-interaction terms.
%}

%%%%%%%%%%%%%%%%%%%%%%%%%%%%%%%%%%%%%%%%%%%%%%%%%%%%%%%%%%
\section{Conclusions and outlook}\label{sec:conclusions}
\noindent
%%%%%%%%%%%%%%%%%%%%%%%%%%%%%%%%%%%%%%%%%%%%%%%%%%%%%%%%%%
We outlined the proposal for a novel geometric phase approach to confinement in Yang--Mills theories. The crucial aspects of the framework we have developed can be summarized as follows:
\begin{itemize}
    \item i) hadrons' ground-states are interpreted geometrically as manifolds. They can be cut away from the vacuum space manifold, since they represent its weakly-interacting sub-manifolds;
    \item ii) the gravitational interaction with the Yang--Mills fields induces stochastic fluctuations of the topology of the sub-manifolds that represent states of matter;
    \item iii) in the dual picture, changes of topology corresponds to the generation of gauge vortices, which finally source the geometric phase of hadrons' ground states.
\end{itemize}
This mechanism ultimately provides the area law for quarks' confinement according to the center vortex mechanism. Nonetheless, several relevant points still need to be addressed.\\

First of all, the formation at the Fermi scale of a non-vanishing condensate is eminently related to the topological properties of the manifold. The energy scale that defines the size of the ground states is still related to dimensional transmutation. The energy scale of confinement is related in this framework to the amplitude of the stochastic noises that enter the stochastic Einstein--Yang--Mills flow. In \cite{lulli2022stochasticquantizationgeneralrelativity} the amplitude of the gravitational stochastic noise was traced back to the variance of its fluctuations, and shown to be proportional to the cosmological constant. On the other hand, for gauge fields it is natural to relate the origin of the amplitude of the stochastic noise to chaos effects in the dynamics of the chromo-electric and chromo-magnetic fluxes, via the Asonov maps \cite{asselmeyermaluga2016smoothquantumgravityexotic}.\\ 

A fascinating possibility, which nonetheless has still to be inspected in detail, is that the stochastic noise of gauge fields could be still decomposed in terms of the gravitational stochastic noise, and finally related to the unbalance of the energy densities of the chromo-electric and chromo-magnetic condensates, as inspected in \cite{Addazi:2018fyo} at the cosmological level. Thus, within this picture, the energy scale of the condensate would be still related to the variance of the stochastic fluctuations of the metric tensor. Outside the Fermi world, this acquires the value of the cosmological constant \cite{lulli2022stochasticquantizationgeneralrelativity}, but when distributions of energy density are taken into account inside the Fermi scale, its value would increase from the one proper of the cosmological constant to that one of $\Lambda_{\rm QCD}$.\\

Dimensional transmutation has been reconsidered in the framework we proposed. We accounted indeed for the breakdown of the conformal symmetry at the quantum level, as being induced by the stochastic gauge-geometry flow, while the system is out of equilibrium. 
The stochastic time regulates the out-of-equilibrium dynamics, making evident that the breakdown of the conformal symmetry (at the quantum level) emerges at non-zero stochastic time distances, and out of equilibrium. Scales can be then individuated in the theory through the coupling between gravity and matter. It is indeed the Newton constant $G$ that provides a dimension-full constant to introduce new physical scale. In other words, it is the interplay between the coupling constant of the theory of gravity and the specific coupling constant of matter that generate the physical relevant scale for the QFT under investigation. \\

Two specific examples of emergence of physical scales through dimensional transmutation --- induced by the coupling with matter, and reminiscent of the conformal anomaly as sourced by the trace of the energy-momentum tensor --- are provided by the theory of QED (the Bohr radius of the electron, which regulates the ``size'' of the Hydrogen atom) and the theory of QCD (the $\Lambda_{\rm QCD}$ scale). Therefore, we may argue that also in the former case the physical scale should be provided by the running  of the values of $G$ that is induced by the stochastic gauge-geometry flow.\\ 

%, together with a rescaling with respect to $g_{\rm em}$ and $g_{\rm s}$ of the gravitational coupling constant $G$. Hence, it would be tempting to consider
%\begin{equation}
%    r_{\rm Bohr}^2 \sim G/g_{\rm em}\,, \qquad 
%    \delta^{\rm QCD}\Lambda  \sim g_{\rm s}/G \,.
%\end{equation}
%having denoted with $\delta^{\rm QCD}\Lambda$ the contribution from the QCD sector to the cosmological constant $\Lambda$.\\

The so-called ``Vacuum Catastrophe'' reflects the fundamental fact that the energy density of the quark-gluon condensate responsible for the color confinement in QCD is off by over forty orders of magnitude and has a wrong sign compared to the observable cosmological $\Lambda$-term value. In the Yang--Mills effective action approach \cite{Savvidy:1977as,Matinyan:1976mp} it is possible to show that the exact compensation of the electric (metastable) and magnetic (stable) gluon condensate components --- localized into vacuum domains spatially separated by domain walls in the QCD vacuum (averaged over space-time volumes above the Fermi scale) --- is the necessary and sufficient condition for confinement in QCD \cite{Addazi:2018fyo,Addazi:2018ctp}. Provided that the energy scales of ``electric gluon'' and ``magnetic gluon'' condensates are not the same, the latter turn out to be formed at different space-time separations. However, they evolve towards the same values of the energy density, but with opposite signs, due to their cosmological attractor nature. This effectively causes the cancellation of ``electric gluon'' and ``magnetic gluon'' contributions to the QCD ground state in macroscopic volumes, i.e.~in the deep infrared limit of the theory \cite{Pasechnik:2013sga,Addazi:2022whi}. The gravitational interactions then break the discrete mirror symmetry of the QCD vacuum already in the semi-classical approximation. Such a breaking causes a non-zeroth (positive) gravitational correction to the resulting effective QCD-induced cosmological constant $\delta^{\rm QCD}\Lambda$, found to be close (within an order of magnitude) to the observed value \cite{Pasechnik:2013poa}.\\

Such a cancellation emerges in experiment as a complete disappearance of the colored degrees of freedom from the theory in the IR limit, i.e. with characteristic wave lengths beyond the Fermi scale, as they are ``trapped'' within the domains at the QCD length scale. This is fully compatible with the classical limit of the Yang--Mills theory, in which the conformal anomaly dynamically vanishes and only a matter-like (non-relativistic) QCD medium (hadron gas) remains at large separations. For the exact cancellation, and hence for color confinement, the restoration of a discrete (mirror) symmetry between the ``electric gluon'' and ``magnetic gluon'' contributions at the level of the effective Lagrangian is an intrinsic property of pure Yang--Mills theory and the RG flow equations.

%\AM{At the next stage, it would be mandatory to interconnect between the effective Yang--Mills action approach, featuring the dimensional transmutation and an emergence of the mirror symmetry of the Yang-Mills vacua, the respective RG flow and the RF formalism [I think this sentence is more for us, so it can be removed].}

\section*{Acknowledgments} 
\noindent
We are indebted with Matteo Lulli for continuous inspiring discussions, especially during the early stage of development of the analysis.

\appendix

%%%%%%%%%%%%%%%%%%%%%%%%%%%%%%%%%%%%%%%%%%%%%%%%%%%%%%%%%%
\section{$3$-Manifold topology and geometry}\label{sec:topology}
\noindent 
%%%%%%%%%%%%%%%%%%%%%%%%%%%%%%%%%%%%%%%%%%%%%%%%%%%%%%%%%%
We hereby collect some general facts on $3$-manifolds and their geometry.  A connected 3-manifold $N$ is prime if it cannot be obtained as a connected sum of two manifolds $N_{1}\#N_{2}$, neither of which is the 3-sphere $S^{3}$ --- or, equivalently, neither of which is homeomorphic to $N$. Examples are the 3-torus $T^{3}$ and $S^{1}\times S^{2}$, as well as the Poincar\'e sphere. According to \cite{Milnor1962ADT}, any compact, oriented 3-manifold is the connected sum of a unique (up to homeomorphism) collection of prime 3-manifolds (prime decomposition). A subset of prime manifolds are the irreducible 3-manifolds. A connected 3-manifold is irreducible if every differentiable submanifold $S$ homeomorphic to a sphere $S^{2}$ bounds a subset $D$ (i.e.~$\partial D=S$) that is homeomorphic to the closed ball $D^{3}$. Moreover, it is known that the only (orientable) prime but reducible 3-manifold is $S^{1}\times S^{2}$. \\

To understand the geometric properties, a finer decomposition induced by incompressible tori is needed. A properly embedded connected surface $S\subset N$ is called 2-sided if its normal bundle is trivial, and 1-sided if its normal bundle is nontrivial. A 2-sided connected surface $S$ other than $S^{2}$ or $D^{2}$ is called incompressible if for each disk $D\subset N$ with $D\cap S=\partial D$ there is a disk $D'\subset S$ with $\partial D'=\partial D$. The boundary of a 3-manifold is an incompressible surface. Most importantly, the 3-sphere $S^{3}$, $S^{2}\times S^{1}$ and the 3-manifolds $S^{3}/\Gamma$ with $\Gamma\subset SO(4)$ a finite subgroup do not contain incompressible surfaces. The class of 3-manifolds $S^{3}/\Gamma$ (the spherical 3-manifolds) include cases like either the Poincar\'e sphere ($\Gamma=I^{*}$ the binary icosahedral group) or lens spaces ($\Gamma=\mathbb{Z}_{p}$ the cyclic group). \\

Let $K_{i}$ be irreducible 3-manifolds containing incompressible surfaces, and assume that $N$ splits into pieces (along embedded $S^{2}$) 
\begin{eqnarray}\label{eq:prime-decomposition}
\begin{aligned}
N&=&K_{1}\#\cdots\#K_{n_{1}}\#_{n_{2}}S^{1}\times S^{2}\#_{n_{3}}S^{3}/\Gamma\,,
\end{aligned}
\end{eqnarray}
where $\#_{n}$ denotes the $n$-fold connected sum and $\Gamma\subset SO(4)$ is a finite subgroup. The decomposition of $N$ is unique up to the order of the factors. The irreducible 3-manifolds $K_{1},\ldots,\,K_{n_{1}}$ are able to contain incompressible tori. So one can split $K_{i}$ along the tori into simpler pieces $K=H\cup_{T^{2}}G$ \cite{jaco1979seifert} (called the JSJ decomposition). The two classes $G$ and $H$ are the graph manifold $G$ and the hyperbolic 3-manifold $H$.\\

The hyperbolic 3-manifold $H$ has a torus boundary $T^{2}=\partial H$, i.e.~$H$ admits a hyperbolic structure in the interior only. It is also useful to consider the splitting of the link/knot complement. As shown in \cite{budney2005jsj}, the Whitehead double of a knot leads to the JSJ decomposition of the complement into the knot complement and the complement of the Whitehead link (along one torus boundary of the Whitehead link complement).\\

One property of hyperbolic 3-manifolds is central for our considerations: the Mostow rigidity. As shown by Mostow \cite{Mos:68}, every hyperbolic $n-$manifold $n>2$ with finite volume has the property that every diffeomorphism (especially, every conformal transformation) is induced by an isometry. Therefore, one cannot scale a hyperbolic 3-manifold, and thus the volume is a topological invariant. Together with the prime and the JSJ decomposition, 
\begin{eqnarray}
\begin{aligned}
N&=&\left(H_{1}\cup_{T^{2}}G_{1}\right)\#\cdots\#\left(H_{n_{1}}\cup_{T^{2}}G_{n_{1}}\right)\\
    &&\#_{n_{2}}S^{1}\times S^{2}\#_{n_{3}}S^{3}/\Gamma \,,
\end{aligned}
\end{eqnarray}
we may then discuss the geometric properties central to the Thurstons geometrization theorem: every oriented closed prime 3-manifold can be cut along tori (JSJ decomposition), so that the interior of each of the resulting manifolds has a geometric structure with finite volume. We hereby clarify the term `geometric structure'. A model geometry is a simply connected smooth manifold $X$ together with a transitive action of a Lie group $G$ on $X$ with compact stabilizers. \\

A geometric structure on a manifold $N$ is a diffeomorphism from $N$ to $X/\Gamma$ for some model geometry $X$, where $\Gamma$ is a discrete subgroup of $G$ acting freely on $X$. It is a surprising fact that there are also a finite number of three-dimensional model geometries, i.e.~only eight different geometries with the following structure: spherical $(S^{3},O_{4}(\mathbb{R}))$, Euclidean $(\mathbb{E}^{3},O_{3}(\mathbb{R})\ltimes\mathbb{R}^{3})$, hyperbolic $(\mathbb{H}^{3},O_{1,3}(\mathbb{R})^{+})$, mixed spherical-Euclidean $(S^{2}\times\mathbb{R},O_{3}(\mathbb{R})\times\mathbb{R}\times\mathbb{Z}_{2})$, mixed hyperbolic-Euclidean $(\mathbb{H}^{2}\times\mathbb{R},O_{1,3}(\mathbb{R})^{+}\times\mathbb{R}\times\mathbb{Z}_{2})$ and 3 exceptional cases called $\tilde{SL}_{2}$ (twisted version of $\mathbb{H}^{2}\times\mathbb{R}$), NIL (geometry of the Heisenberg group as twisted version of $\mathbb{E}^{3}$), SOL (split extension of $\mathbb{R}^{2}$ by $\mathbb{R}$, i.e.~the Lie algebra of the group of isometries of 2-dimensional Minkowski space).

\section{Embedding Casson handle by using quadratic differentials and vertical foliations}\label{Casson}
\noindent

A rigorous mathematical argumentation is based on the embedding of the tree $T_{CH}$ for a Casson handle $CH$. The Casson handle is a branched surface and it can be described by quadratic differentials as follows. 
Let $X$ be a Riemannian surface, then a quadratic differential is a section of $T^{*}X^{1,0}\otimes T^{*}X^{1,0}$, which is locally given by 
\[
q=q(z)dz^{2}=q(z)\,dz\otimes dz\,,
\]
with $q(z)$ holomorphic function. Away from the zeros of $q(z)$
we can choose a canonical conformal coordinate $\xi(z)=\intop^{z}\sqrt{q}$
so that $q=d\xi^{2}$. Then the set $\left\{ Re(\xi)=const.\right\} $
defines a foliation, called the vertical measured foliation. The holomorphic
function $q(z)$ can be locally expressed as a polynomial. The zeros
of the polynomial are the branching points of the surface, i.e. $z^{n}$
branches into $n$ pieces at $z=0$. Using this result, we are able
to generate the tree $T_{CH}$ by a polynomial. Furthermore we identify
the coordinate $t$ with $t=Re(\xi)$ so that $t=const.$ defines
a vertical foliation (into slices of the constant time). By the theorem
of Hubbard and Masur \cite{HubbardMasur:1979}, for every measured
foliation on $X$ (of genus $g>1$) there exists a unique quadratic
differential so that its vertical measured foliation is equivalent
to the measured foliation. In our case, the infinite tree seen as
branched surface is the covering space of a Riemannian surface of
infinite genus. Therefore the quadratic differential is unique by
lifting it to the covering.

\vspace{2cm}
\bibliography{references}

@article{Ito1951, title={On a Formula Concerning Stochastic Differentials}, volume={3}, DOI={10.1017/S0027763000012216}, journal={Nagoya Mathematical Journal}, author={Itô, Kiyosi}, year={1951}, pages={55–65}}

@article{Greensite:2016pfc,
    author = "Greensite, Jeff",
    editor = "Foka, Y. and Brambilla, N. and Kovalenko, V.",
    title = "{Confinement from Center Vortices: A review of old and new results}",
    eprint = "1610.06221",
    archivePrefix = "arXiv",
    primaryClass = "hep-lat",
    doi = "10.1051/epjconf/201713701009",
    journal = "EPJ Web Conf.",
    volume = "137",
    pages = "01009",
    year = "2017"
}

@article{Engelhardt:1998wu,
    author = "Engelhardt, M. and Langfeld, K. and Reinhardt, H. and Tennert, O.",
    title = "{Interaction of confining vortices in SU(2) lattice gauge theory}",
    eprint = "hep-lat/9801030",
    archivePrefix = "arXiv",
    reportNumber = "UNITU-THEP-4-98",
    doi = "10.1016/S0370-2693(98)00583-8",
    journal = "Phys. Lett. B",
    volume = "431",
    pages = "141--146",
    year = "1998"
}

@article{Pasechnik:2021ncb,
    author = "Pasechnik, Roman and \v{S}umbera, Michal",
    title = "{Different Faces of Confinement}",
    eprint = "2109.07600",
    archivePrefix = "arXiv",
    primaryClass = "hep-ph",
    doi = "10.3390/universe7090330",
    journal = "Universe",
    volume = "7",
    number = "9",
    pages = "330",
    year = "2021"
}

@article{Addazi:2018ctp,
    author = "Addazi, Andrea and Marcian\`o, Antonino and Pasechnik, Roman",
    title = "{Time-crystal ground state and production of gravitational waves from QCD phase transition}",
    eprint = "1812.07376",
    archivePrefix = "arXiv",
    primaryClass = "hep-th",
    doi = "10.1088/1674-1137/43/6/065101",
    journal = "Chin. Phys. C",
    volume = "43",
    number = "6",
    pages = "065101",
    year = "2019"
}

@article{Addazi:2022whi,
    author = {Addazi, Andrea and Lundberg, Torbj\"orn and Marcian\`o, Antonino and Pasechnik, Roman and \v{S}umbera, Michal},
    title = "{Cosmology from Strong Interactions}",
    eprint = "2204.02950",
    archivePrefix = "arXiv",
    primaryClass = "hep-ph",
    doi = "10.3390/universe8090451",
    journal = "Universe",
    volume = "8",
    number = "9",
    pages = "451",
    year = "2022"
}

@article{Mandelstam:1974pi,
    author = "Mandelstam, S.",
    title = "{Vortices and Quark Confinement in Nonabelian Gauge Theories}",
    reportNumber = "PRINT-74-1623 (UC,BERKELEY)",
    doi = "10.1016/0370-1573(76)90043-0",
    journal = "Phys. Rept.",
    volume = "23",
    pages = "245--249",
    year = "1976"
}

@proceedings{Zichichi:1975izv,
    author = "'t Hooft, G.",
    editor = "Zichichi, A.",
    title = "{High-Energy Physics: Proceedings, EPS International Conference, Palermo, Italy, 23-28 June 1975.}",
    publisher = "Compositori",
    address = "Bologna, Italy",
    month = "6",
    year = "1975"
}

@article{Polyakov:1975rs,
    author = "Polyakov, Alexander M.",
    editor = "Taylor, J. C.",
    title = "{Compact Gauge Fields and the Infrared Catastrophe}",
    doi = "10.1016/0370-2693(75)90162-8",
    journal = "Phys. Lett. B",
    volume = "59",
    pages = "82--84",
    year = "1975"
}

@article{tHooft:1977nqb,
    author = "'t Hooft, Gerard",
    title = "{On the Phase Transition Towards Permanent Quark Confinement}",
    reportNumber = "Print-78-0099 (UTRECHT)",
    doi = "10.1016/0550-3213(78)90153-0",
    journal = "Nucl. Phys. B",
    volume = "138",
    pages = "1--25",
    year = "1978"
}

@article{Diamantini:2018mjg,
    author = "Diamantini, M. C. and Trugenberger, C. A. and Vinokur, V. M.",
    title = "{Confinement and Asymptotic Freedom with Cooper pairs}",
    eprint = "1807.01984",
    archivePrefix = "arXiv",
    primaryClass = "hep-th",
    journal = "APS Physics",
    volume = "1",
    pages = "77",
    year = "2018"
}

@misc{lulli2022stochasticquantizationgeneralrelativity,
      title={Stochastic Quantization of General Relativity \`a la Ricci-Flow}, 
      author={Matteo Lulli and Antonino Marciano and Xiaowen Shan},
      year={2025},
      journal = "Fortschritte der Physik",
      eprint={2112.01490},
      archivePrefix={arXiv},
      primaryClass={gr-qc},
      url={https://onlinelibrary.wiley.com/doi/10.1002/prop.70041}, 
}

@article{Matinyan:1976mp,
    author = "Matinyan, Sergei G. and Savvidy, G. K.",
    title = "{Vacuum Polarization Induced by the Intense Gauge Field}",
    reportNumber = "EFI-172-18-76-YEREVAN",
    doi = "10.1016/0550-3213(78)90463-7",
    journal = "Nucl. Phys. B",
    volume = "134",
    pages = "539--545",
    year = "1978"
}

@article{Pasechnik:2013sga,
    author = "Pasechnik, Roman and Beylin, Vitaly and Vereshkov, Grigory",
    title = "{Possible compensation of the QCD vacuum contribution to the dark energy}",
    eprint = "1302.5934",
    archivePrefix = "arXiv",
    primaryClass = "gr-qc",
    doi = "10.1103/PhysRevD.88.023509",
    journal = "Phys. Rev. D",
    volume = "88",
    number = "2",
    pages = "023509",
    year = "2013"
}

@article{Pasechnik:2013poa,
    author = "Pasechnik, Roman and Beylin, Vitaly and Vereshkov, Grigory",
    title = "{Dark Energy from graviton-mediated interactions  in the QCD vacuum}",
    eprint = "1302.6456",
    archivePrefix = "arXiv",
    primaryClass = "gr-qc",
    doi = "10.1088/1475-7516/2013/06/011",
    journal = "JCAP",
    volume = "06",
    pages = "011",
    year = "2013"
}

@article{Addazi:2018fyo,
    author = "Addazi, Andrea and Marcian\`o, Antonino and Pasechnik, Roman and Prokhorov, George",
    title = "{Mirror Symmetry of quantum Yang-Mills vacua and cosmological implications}",
    eprint = "1804.09826",
    archivePrefix = "arXiv",
    primaryClass = "hep-th",
    doi = "10.1140/epjc/s10052-019-6780-x",
    journal = "Eur. Phys. J. C",
    volume = "79",
    number = "3",
    pages = "251",
    year = "2019"
}

@article{Savvidy:1977as,
    author = "Savvidy, G. K.",
    title = "{Infrared Instability of the Vacuum State of Gauge Theories and Asymptotic Freedom}",
    reportNumber = "EFI-214-6-77-YEREVAN",
    doi = "10.1016/0370-2693(77)90759-6",
    journal = "Phys. Lett. B",
    volume = "71",
    pages = "133--134",
    year = "1977"
}

@article{Parisi:1980ys,
    author = "Parisi, G. and Wu, Yong-shi",
    title = "{Perturbation Theory Without Gauge Fixing}",
    reportNumber = "ASITP-80-004",
    journal = "Sci. Sin.",
    volume = "24",
    pages = "483",
    year = "1981"
}

@article{Lulli:2023fcl,
    author = "Lulli, Matteo and Marciano, Antonino and Piscicchia, Kristian",
    title = "{Stochastic Ricci Flow dynamics of the gravitationally induced wave-function collapse}",
    eprint = "2307.10136",
    archivePrefix = "arXiv",
    primaryClass = "gr-qc",
    month = "7",
    year = "2023"
}

@article{cotta1990quantum,
  title={Quantum field theory and link invariants},
  author={Cotta-Ramusino, P and Guadagnini, Enore and Martellini, M and Mintchev, M},
  journal={Nuclear Physics B},
  volume={330},
  number={2-3},
  pages={557--574},
  year={1990},
  publisher={Elsevier}
}

@article{Rumpf:1985eh,
    author = "Rumpf, Helmut",
    title = "{Stochastic Quantization of Einstein Gravity}",
    reportNumber = "UWThPh-1985-15",
    doi = "10.1103/PhysRevD.33.942",
    journal = "Phys. Rev. D",
    volume = "33",
    pages = "942",
    year = "1986"
}

@article{Elitzur:1975im,
    author = "Elitzur, S.",
    title = "{Impossibility of Spontaneously Breaking Local Symmetries}",
    doi = "10.1103/PhysRevD.12.3978",
    journal = "Phys. Rev. D",
    volume = "12",
    pages = "3978--3982",
    year = "1975"
}

@article{Kugo:1979gm,
    author = "Kugo, Taichiro and Ojima, Izumi",
    title = "{Local Covariant Operator Formalism of Nonabelian Gauge Theories and Quark Confinement Problem}",
    reportNumber = "KUNS-493",
    doi = "10.1143/PTPS.66.1",
    journal = "Prog. Theor. Phys. Suppl.",
    volume = "66",
    pages = "1--130",
    year = "1979"
}

@inproceedings{Kugo:1995km,
    author = "Kugo, Taichiro",
    title = "{The Universal renormalization factors Z(1) / Z(3) and color confinement condition in nonAbelian gauge theory}",
    booktitle = "{International Symposium on BRS Symmetry on the Occasion of Its 20th Anniversary}",
    eprint = "hep-th/9511033",
    archivePrefix = "arXiv",
    reportNumber = "KUNS-1368",
    pages = "107--119",
    month = "7",
    year = "1995"
}

@article{Hata:1981nd,
    author = "Hata, Hiroyuki",
    title = "{Restoration of the Local Gauge Symmetry and Color Confinement in Nonabelian Gauge Theories}",
    reportNumber = "KUNS 601",
    doi = "10.1143/PTP.67.1607",
    journal = "Prog. Theor. Phys.",
    volume = "67",
    pages = "1607",
    year = "1982"
}

@article{Hata:1983cs,
    author = "Hata, H.",
    title = "{RESTORATION OF THE LOCAL GAUGE SYMMETRY AND COLOR CONFINEMENT IN NONABELIAN GAUGE THEORIES. II}",
    doi = "10.1143/PTP.69.1524",
    journal = "Prog. Theor. Phys.",
    volume = "69",
    pages = "1524--1536",
    year = "1983"
}

@incollection{jaco1979seifert,
  title={Seifert fibered spaces in 3-manifolds},
  author={Jaco, William and Shalen, Peter B},
  booktitle={Geometric topology},
  pages={91--99},
  year={1979},
  publisher={Elsevier}
}

@article{budney2005jsj,
  title={JSJ-decompositions of knot and link complements in the 3-sphere},
  author={Budney, Ryan},
  journal={arXiv preprint math/0506523},
  year={2005}
}

@article{Mos:68,
     author = {Mostow, G. D.},
     title = {Quasi-conformal mappings in $n$-space and the rigidity of hyperbolic space forms},
     journal = {Publications Math\'ematiques de l'IH\'ES},
     pages = {53--104},
     publisher = {Institut des Hautes \'Etudes Scientifiques},
     volume = {34},
     year = {1968},
     mrnumber = {38 #4679},
     zbl = {0189.09402},
     language = {en},
     url = {http://www.numdam.org/item/PMIHES_1968__34__53_0/}
}

@book{Greensite:2011zz,
    author = "Greensite, Jeff",
    title = "{An introduction to the confinement problem}",
    doi = "10.1007/978-3-642-14382-3",
    volume = "821",
    year = "2011"
}

@article{Halpern:1978ik,
    author = "Halpern, M. B.",
    title = "{Field Strength and Dual Variable Formulations of Gauge Theory}",
    reportNumber = "LBL-7980",
    doi = "10.1103/PhysRevD.19.517",
    journal = "Phys. Rev. D",
    volume = "19",
    pages = "517",
    year = "1979"
}

@article{Batrouni:1984rb,
    author = "Batrouni, Ghassan G. and Halpern, M. B.",
    title = "{String, Corner and Plaquette Formulation of Finite Lattice Gauge Theory}",
    reportNumber = "CLNS-84/602",
    doi = "10.1103/PhysRevD.30.1782",
    journal = "Phys. Rev. D",
    volume = "30",
    pages = "1782",
    year = "1984"
}

@article{Intriligator:1995er,
    author = "Intriligator, Kenneth A. and Seiberg, N.",
    editor = "Bars, I. and Bouwknegt, P. and Minahan, J. and Nemeschansky, D. and Pilch, K. and Saleur, H. and Warner, N. P.",
    title = "{Phases of N=1 supersymmetric gauge theories and electric - magnetic triality}",
    eprint = "hep-th/9506084",
    archivePrefix = "arXiv",
    reportNumber = "RU-95-40, IASSNS-HEP-95-48",
    journal = "Nucl. Phys. B Proc. Suppl.",
    volume = "39",
    pages = "1",
    year = "1996"
}

@article{DiGiacomo:2000irz,
    author = "Di Giacomo, A. and Dosch, Hans Gunter and Shevchenko, V. I. and Simonov, Yu. A.",
    title = "{Field correlators in QCD: Theory and applications}",
    eprint = "hep-ph/0007223",
    archivePrefix = "arXiv",
    doi = "10.1016/S0370-1573(02)00140-0",
    journal = "Phys. Rept.",
    volume = "372",
    pages = "319--368",
    year = "2002"
}

@article{Kuzmenko:2004hk,
    author = "Kuzmenko, D. S. and Shevchenko, V. I. and Simonov, Yu. A.",
    title = "{The QCD vacuum, confinement and strings in the vacuum correlator method}",
    eprint = "hep-ph/0310190",
    archivePrefix = "arXiv",
    doi = "10.1070/PU2004v047n01ABEH001696",
    journal = "Phys. Usp.",
    volume = "47",
    pages = "1--15",
    year = "2004"
}

@article{Kondo:1999tj,
    author = "Kondo, Kei-Ichi and Taira, Yutaro",
    title = "{NonAbelian Stokes theorem and quark confinement in SU(N) Yang-Mills gauge theory}",
    eprint = "hep-th/9911242",
    archivePrefix = "arXiv",
    reportNumber = "CHIBA-EP-117",
    doi = "10.1143/PTP.104.1189",
    journal = "Prog. Theor. Phys.",
    volume = "104",
    pages = "1189--1265",
    year = "2000"
}

@article{Kondo:2000pp,
    author = "Kondo, K. I. and Taira, Y.",
    title = "{NonAbelian Stokes Theorem and Quark confinement in SU(3) Yang-Mills gauge theory}",
    eprint = "hep-th/9906129",
    archivePrefix = "arXiv",
    reportNumber = "CHIBA-EP-114",
    doi = "10.1142/S0217732300000359",
    journal = "Mod. Phys. Lett. A",
    volume = "15",
    pages = "367--377",
    year = "2000"
}

@article{Diakonov:2000kw,
    author = "Diakonov, Dmitri and Petrov, Victor",
    title = "{NonAbelian Stokes theorems in Yang-Mills and gravity theories}",
    eprint = "hep-th/0008035",
    archivePrefix = "arXiv",
    reportNumber = "NORDITA-2000-65-HE",
    doi = "10.1134/1.1385630",
    journal = "J. Exp. Theor. Phys.",
    volume = "92",
    pages = "905--920",
    year = "2001"
}

@article{Hirayama:1999ar,
    author = "Hirayama, Minoru and Ueno, Masataka",
    title = "{NonAbelian Stokes theorem for Wilson loops associated with general gauge groups}",
    eprint = "hep-th/9907063",
    archivePrefix = "arXiv",
    doi = "10.1143/PTP.103.151",
    journal = "Prog. Theor. Phys.",
    volume = "103",
    pages = "151--159",
    year = "2000"
}

@article{Karp:1999vq,
    author = "Karp, Robert L. and Mansouri, Freydoon and Rno, Jung S.",
    title = "{Product integral formalism and nonAbelian Stokes theorem}",
    eprint = "hep-th/9910173",
    archivePrefix = "arXiv",
    reportNumber = "UCTP-117-99",
    doi = "10.1063/1.533068",
    journal = "J. Math. Phys.",
    volume = "40",
    pages = "6033--6043",
    year = "1999"
}

@article{Diakonov:1989fc,
    author = "Diakonov, Dmitri and Petrov, V. Yu.",
    title = "{A Formula for the Wilson Loop}",
    doi = "10.1016/0370-2693(89)91062-9",
    journal = "Phys. Lett. B",
    volume = "224",
    pages = "131--135",
    year = "1989"
}

@article{Fishbane:1980eq,
    author = "Fishbane, Paul M. and Gasiorowicz, Stephen and Kaus, Peter",
    title = "{Stokes' Theorems for Nonabelian Fields}",
    reportNumber = "Print-80-0701 (MINNESOTA), COO-1764-405",
    doi = "10.1103/PhysRevD.24.2324",
    journal = "Phys. Rev. D",
    volume = "24",
    pages = "2324",
    year = "1981"
}

@article{Arefeva:1979dp,
    author = "Arefeva, Irina",
    title = "{NonAbelian Stokes formula}",
    doi = "10.1007/BF01018469",
    journal = "Theor. Math. Phys.",
    volume = "43",
    pages = "353",
    year = "1980"
}

@article{Milnor1962ADT,
  title={A DUALITY THEOREM FOR REIDEMEISTER TORSION},
  author={John W. Milnor},
  journal={Annals of Mathematics},
  year={1962},
  volume={76},
  pages={137},
  url={https://api.semanticscholar.org/CorpusID:121340632}
}

@article{Engelhardt_1998,
   title={Interaction of confining vortices in SU(2) lattice gauge theory},
   volume={431},
   ISSN={0370-2693},
   url={http://dx.doi.org/10.1016/S0370-2693(98)00583-8},
   DOI={10.1016/s0370-2693(98)00583-8},
   number={1–2},
   journal={Physics Letters B},
   publisher={Elsevier BV},
   author={Engelhardt, M. and Langfeld, K. and Reinhardt, H. and Tennert, O.},
   year={1998},
   month=jul, pages={141–146} }

@article{Greensite_2017,
   title={Confinement from Center Vortices: A review of old and new results},
   volume={137},
   ISSN={2100-014X},
   url={http://dx.doi.org/10.1051/epjconf/201713701009},
   DOI={10.1051/epjconf/201713701009},
   journal={EPJ Web of Conferences},
   publisher={EDP Sciences},
   author={Greensite, Jeff},
   editor={Foka, Y. and Brambilla, N. and Kovalenko, V.},
   year={2017},
   pages={01009} }

@article{Myers1983, author = {Myers,T.}, year =1983,
title = {Homology cobordisms, link concordances, and hyperbolic 3-manifolds}, journal ={Trans.of AMS}, volume=278, 
pages ={271-288}
}

@article{Wilson:1974sk,
    author = "Wilson, Kenneth G.",
    editor = "Taylor, J. C.",
    title = "{Confinement of Quarks}",
    reportNumber = "CLNS-262",
    doi = "10.1103/PhysRevD.10.2445",
    journal = "Phys. Rev. D",
    volume = "10",
    pages = "2445--2459",
    year = "1974"
}

@article{Lulli_2024B,
   title={The De Turk term and the fate of torsion in stochastic geometry flow},
   journal={in preparation},
   year={2024},
   author={Lulli, M. and Li, J. and Wang, J. and Marcian\`o, A.} 
}

@article{Flo:88,
    author  = {Floer, A.},
    title   = {An instanton Invariant for 3-manifolds},
    journal = {Comm. Math. Phys.},
    pages   = {215-240},
    year    = {1988},
    volume  = {{\bf 118}}
}

@article{Sei:88,
    author  = {Seiberg, N.},
    title   = {Supersymmetry and non-perturbative beta functions},
    journal = {Phys. Lett. B},
    pages   = {75-80},
    year    = {1988},
    volume  = {206}
}

@book{HarLaw:93,
    author  = {F.R. Harvey and H.B. Lawson},
    title   = {A theory of characteristic currents associated to a singular connection},
    year    = {1993},
    publisher = {Soci{\'e}t{\'e} Math{\'e}matique De France},
    edition = {ast{\'e}risque 213}
}

@article{kleiner2008notes,
  title={Notes on Perelman’s papers},
  author={Kleiner, Bruce and Lott, John},
  journal={Geometry \& Topology},
  volume={12},
  number={5},
  pages={2587--2855},
  year={2008},
  publisher={Mathematical Sciences Publishers}
}

@article{perelman2002entropy,
  title={The entropy formula for the Ricci flow and its geometric applications},
  author={Perelman, Grisha},
  journal={arXiv preprint math/0211159},
  year={2002}
}

@article{perelman2003ricci,
  title={Ricci flow with surgery on three-manifolds},
  author={Perelman, Grisha},
  journal={arXiv preprint math/0303109},
  year={2003}
}

@article{Wit:89.2,
    author  = {Witten, E.},
    title   = {2+1 dimensional gravity as an exactly soluble system},
    journal = {Nucl. Phys.},
    pages   = {46-78},
    year    = {1988/89},
    volume  = {{\bf B311}}
}

@article{Wit:89.3,
    author  = {Witten, E.},
    title   = {Topology-changing amplitudes in 2+1 dimensional gravity},
    journal = {Nucl. Phys.},
    pages   = {113-140},
    year    = {1989},
    volume  = {{\bf B323}}
}

@article{Wit:91.2,
    author  = {Witten, E.},
    title   = {Quantization of {C}hern-{S}imons Gauge Theory with Complex Gauge Group},
    journal = {Comm. Math. Phys.},
    pages   = {29-66},
    year    = {1991},
    volume  = {{\bf 137}}
}

@ARTICLE{MacDMan:1977,
  author = {MacDowell, S.W. and Mansouri, F.},
  year = 1977,
  title = {Unified Geometric Theory of Gravity and Supergravity},
  journal = {Phys. Rev. Lett.},
  volume = {{\bf 38}},
  pages = {739-742},
  note = {doi:10.1103/PhysRevLett.38.739}
}

@UNPUBLISHED{EnergyScaleQCD:2004,
  author = {Sekhar Chivukula, R.},
  year = 2004,
  title = {The Origin of Mass in {QCD}},
  note = {arXiv:hep-ph/0411198}
}

@ARTICLE{EnergyScaleQCD:2016,
  author = {Deur, A. and Brodsky, S.J. and de Teramond, G.F.},
  year = 2016,
  title = {The {QCD} Running Coupling},
  journal = {Prog. Part. Nuc. Phys.},
  volume = {{\bf 90}},
  pages = {1-74},
  note = {arXiv:1604.08082}
}

@ARTICLE{HubbardMasur:1979,
  author = {Hubbard, J. and Masur, H.},
  year = 1979,
  title = {Quadratic Differentials and Foliations},
  journal = {Acta Math.},
  volume = {{\bf 142}},
  pages = {221-274}
}

@ARTICLE{AsselmeyerKrol2018a,
  author = {Asselmeyer-Maluga, T. and Krol, J.},
  year = 2018,
  title = {How to Obtain a Cosmological Constant from Small Exotic
          ${\mathbb R}^4$},
  journal = {Physics of the Dark Universe},
  volume = {{\bf 19}},
  pages = {66-77},
  note = {arXiv:1709.03314}
}

@misc{asselmeyermaluga2016smoothquantumgravityexotic,
      title={Smooth quantum gravity: Exotic smoothness and Quantum gravity}, 
      author={Torsten Asselmeyer-Maluga},
      year={2016},
      eprint={1601.06436},
      archivePrefix={arXiv},
      primaryClass={gr-qc},
      url={https://arxiv.org/abs/1601.06436}, 
}

\end{document}